\documentclass[onecolumn,pra,aps,superscriptaddress,english,floatfix,notitlepage]{revtex4-1}
\usepackage{amsmath}
\usepackage{amssymb}
\usepackage{amsfonts}
\usepackage[pdftex]{graphicx}
\usepackage{subfigure} 
\usepackage[english]{babel}
\usepackage{braket}
\usepackage{color}
\usepackage{mathtools}
\usepackage{threeparttable}
\usepackage[colorlinks,linkcolor=blue,anchorcolor=blue,citecolor=blue,urlcolor=blue]{hyperref}
\usepackage{babel}

\makeatother

\usepackage{babel}
\begin{document}

\title{Single-Photon Pulse Induced Transient Entanglement Force}

\author{Li-Ping Yang}
\affiliation{Birck Nanotechnology Center and Purdue Quantum Center, School of Electrical
and Computer Engineering, Purdue University, West Lafayette, IN 47906,
U.S.A.}

\author{Chinmay Khandekar}
\affiliation{Birck Nanotechnology Center and Purdue Quantum Center, School of Electrical
and Computer Engineering, Purdue University, West Lafayette, IN 47906,
U.S.A.}

\author{Tongcang Li}
\affiliation{Department of Physics and Astronomy, Purdue University, West Lafayette, Indiana 47907, USA}
\affiliation{Birck Nanotechnology Center and Purdue Quantum Center, School of Electrical
and Computer Engineering, Purdue University, West Lafayette, IN 47906,
U.S.A.}

\author{Zubin Jacob}
\affiliation{Birck Nanotechnology Center and Purdue Quantum Center, School of Electrical
and Computer Engineering, Purdue University, West Lafayette, IN 47906,
U.S.A.}



\email{zjacob@purdue.edu}

\begin{abstract}
We show that a single photon pulse incident on two interacting two-level atoms induces a transient entanglement force between them. After absorption of a multi-mode Fock state pulse, the time-dependent atomic interaction mediated by the vacuum fluctuations changes from the van der Waals interaction to the resonant dipole-dipole interaction (RDDI). We explicitly show that the RDDI force induced by the single photon pulse fundamentally arises from the two-body transient entanglement between the atoms. This single-photon-pulse-induced entanglement force can be continuously tuned from being repulsive to attractive by varying the polarization of the pulse. We further demonstrate that the entanglement force can be enhanced by more than three orders of magnitude if the atomic interactions are mediated by graphene plasmons. These results demonstrate the potential of shaped single photon pulses as a powerful tool to manipulate this entanglement force and also provide a new approach to witness transient atom-atom entanglement. 
\end{abstract}

\maketitle

\section{Introduction}

Single-photon-induced forces and torques correspond to the fundamental limit of optical linear momentum and angular momentum exchange with atoms \cite{tang2014prospect}. Their direct detection is an open challenge since state-of-the-art quantum detectors are only sensitive to energy and arrival time of single photons~\cite{eisaman2011invited}. Recent advances in temporal shaping of single photon scattering from atoms has shed light on the role of the temporal waveform of Fock states~\cite{Romero2012ultrfast}. In light of these developments, it is an open question how single photon waveforms influence dipole-dipole interactions between atoms. Of particular interest is the exploration whether single photon shaped waveforms incident on interacting atoms can lead to experimentally observable transient effects.

During the last two decades, many techniques have been utilized to enhance the strength of the dipole-dipole interaction and the corresponding force~\cite{buhmann2007dispersion}, such as utilizing micro-cavity~\cite{Agarwal1998microcavity,Hopmeier1999enhanced,Donaire2017dipole,cortes2017fundamental}, surface plasmons~\cite{gonzalez2011entanglement,Xu2011entanglement,Haakh2014dynamical}, and hyperbolic materials~\cite{cortes2017super}. Especially, the strong dipole-dipole interaction induced large energy shift in highly excited atoms (e.g. Rydberg atoms) has been proposed as the mechanism for 
``Rydberg blockade", which provides a novel approach for quantum information processing~\cite{Jaksch2000fast,Lukin2001dipole} and simulation of quantum phase transition~\cite{labuhn2016tunable,bernien2017probing}. However, single-photon pulse as a tool to manipulate the transient dipole-dipole force has not been explored.

In this paper, we show the existence of a unique transient entanglement force between two neutral atoms induced by a single photon pulse. With the help of our defined force operator, we explicitly show that the resonant dipole-dipole interaction (RDDI) force fundamentally arises from two-body entanglement, which is significantly different from the van der Waals force. Our theoretical framework combines quantum theories of single-photon pulse scattering~\cite{Domokos2002quantum,wang2011efficient,Baragiola2012n-photon,yang2018concept} and the macroscopic quantum electrodynamics approach of dipole-dipole interaction~\cite{craig1998molecular,Dung2002resonant,Safari2006body,shengwen2018magnetic}. We thus show that the quantum statistics of the incident (Fock-state vs coherent-state) pulses lead to significant differences in the induced RDDI entanglement forces. After absorption of a single photon pulse, the inter-atomic force changes from the extremely weak van der Waals force~\cite{london1937general,Casimir1948retardation,buhmann2007dispersion} to the RDDI force~\cite{mclone1964interaction,stephen1964first} with the amplitude enhanced by $\sim 10$ orders of magnitude. 

\begin{figure*}
\centering
\includegraphics[width=15.5cm]{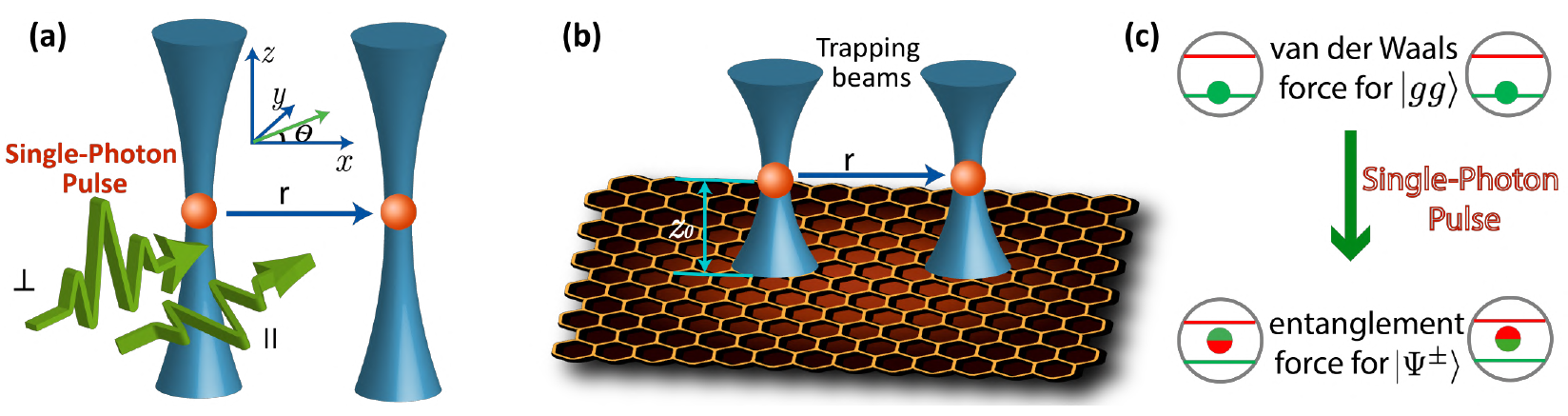}
\caption{\label{fig:1}Schematic of the single-photon pulse induced entanglement force detection. (a) Two atoms in free space. (b) Two atoms on top of a graphene layer ($z_0$ is the height). These two atoms (the yellow spheres) are levitated by two separated optical tweezers. The relative displacement between the two atoms is $\boldsymbol{r}=\boldsymbol{x}_{2}-\boldsymbol{x}_{1}=r\boldsymbol{e}_{x}$, which is along $x$-axis. The linearly polarized single photon pulse propagates along $y$-direction, with polarization being parallel ($\parallel$ with $\theta=0$) or
perpendicular ($\perp$ with $\theta=\pi/2$) to $\boldsymbol{r}$. For two ground-state atoms, the van der Waals force mediated by the vacuum fluctuations is extremely small ($\sim10^{-35}$ N for $r\approx1\ {\rm \mu m}$, far beyond the state-of-art of the force sensitivity~\cite{biercuk2010ultrasensitive,Parazzoli2012observation}). After absorption of a single photon pulse, the atom-atom interaction changes to the RDDI as shown in (c). The corresponding force is enhanced more than $10$ orders to $\sim10^{-21}$ N. We emphasize that this RDDI force for atoms on states $|\Psi^{\pm}\rangle = (|eg\rangle \pm |eg\rangle)/\sqrt{2}$ is an entanglement force, which is fundamentally different from the van der Waals force.}
\end{figure*}

We propose an experiment to detect this single photon pulse induced force with two levitated neutral atoms (see Fig.~\ref{fig:1}), which are separated with distance~$r\sim 1\rm{\mu m}$ by optical tweezers operating at the magic wavelength~\cite{barredo2016atom,endres2016atom,liu2018building}. Even with this enhancement, detection of such a weak transient RDDI force is still a difficult challenge. Therefore, we  we demonstrate that the single photon pulse induced RDDI force can be significantly enhanced by placing the atoms near a graphene layer with the assistance of graphene-based surface-plasmon polaritons.  By investigating the full quantum dynamics of single-photon absorption, we predict optimum entanglement generation mechanisms conducive to experimental inquiry. Finally, we argue that the proposed effect can be differentiated from previously known dipolar interactions since the single photon pulse induced entanglement force can be tuned from repulsive to attractive by tuning the polarization of the incident pulse.

\section{Dipole-dipole interaction force operator}

With the help of the Hellmann-Feynman theorem~\cite{Feynman1937force}, we define a quantum operator to characterize the force generated by the coherent part of the dipole-dipole interaction in~\ref{sec:force_def},
\begin{equation}
\hat{F}(r)\equiv -\frac{\partial}{\partial r}\hat{U}(r) =\sum_{mn}F_{mn}(r)\left|m\right\rangle \left\langle n\right|,\label{eq:force}
\end{equation}
where $F_{mn}(r)\equiv-\partial U_{mn}(r)/\partial r$ is determined by the atom-atom interaction $\hat{U}(r)=\sum_{mn}U_{mn}(r)\left|m\right\rangle \left\langle n\right|$ induced by electromagnetic vacuum fluctuations~\cite{ficek2002entangled,Dung2002resonant} and $\left|m \right\rangle \in\{|gg\rangle ,|eg\rangle ,\left|ge\right\rangle ,\left|ee\right\rangle \}$ for a two-level-atom pair. The dipole-dipole interaction force is always along the axis joining the two atoms. Our defined force operator allows us directly to classify the dipole-dipole interaction force into two categories: (1) van der Waals force between two atoms in a direct-product state, such as the force for two ground-state atoms $\hat{F}_{\rm vdW}=F_{gg,gg}|gg\rangle\langle gg|$; (2) RDDI force for entangled atoms, e.g.,
\begin{equation}
\hat{F}_{\rm RDDI} (r) = F_{eg,ge}(r)\left|eg\right\rangle \left\langle ge\right|+{\rm h.c.}    
\end{equation}
We will show how to control this force with a single photon pulse later.

We emphasize that the latter RDDI force fundamentally arises from two-body entanglement~\cite{Behunin2010non}. The eigenvectors of the force operator $\hat{F}_{\rm RDDI} (r)$ are the two
Bell states
\begin{equation}
\left|\Psi^{\pm}\right\rangle =\frac{1}{\sqrt{2}}\left(\left|eg\right\rangle \pm\left|ge\right\rangle \right),
\end{equation}
with eigenvalues $\pm F_{eg,ge}(r)$. For a given two-atom state $\rho(t)$, the absolute value of the RDDI force is proportional to to the probability difference of the two-atom state on these two entangled states, i.e., $F_{\rm RDDI} (r,t)\propto|\langle\Psi^{+}|\rho(t)|\Psi^{+}\rangle-\langle\Psi^{-}|\rho(t)|\Psi^{-}\rangle|$. This immediately reveals that, to maximize the RDDI force, one needs to prepare the atom pair in one of these two entangled states. We also note that, the RDDI force presents a readout of two-body entanglement. This entanglement force between transition dipoles is fundamentally different from van der Waals force~\cite{Casimir1948retardation} and the force generated by the permanent dipole-dipole interaction~\cite{craig1998molecular}. We emphasize that the maximum possible RDDI force (the eigenvalue of the force operator) is determined by the atom-atom distance $r$. However, the exact time-dependent envelope of the RDDI force in a specific dynamical process is determined by the atomic two-body entanglement. 

\section{Dynamical entanglement force}

The master equation method has been broadly applied to study the dipole-dipole interaction and entanglement between neutral atoms~\cite{ficek2002entangled,Dung2002resonant,kastel2005suppresion,brooke2008super,Dzsotjan2010quantum}. We now incorporate the single photon pulse absorption dynamics with the traditional master equation to show the time-dependent entanglement force induced by a single photon pulse (see~\ref{sec:TDMEQ}), 
\begin{equation}
\frac{d}{dt}\tilde{\rho}(t)=[\hat{\hat{\mathcal{L}}}_{{\rm atom}}+\hat{\hat{\mathcal{L}}}_{{\rm pump}}(t)]\tilde{\rho}(t),\label{eq:TDME1}
\end{equation}
where $\tilde{\rho}(t)=\rho_{{\rm PN}}(t)\otimes\rho(t)$ is an effective density matrix. We have introduced an extra qubit degree of freedom $\rho_{{\rm PN}}(t)$ to characterize the photon number degree (see more details in~\cite{Baragiola2012n-photon}). The initial
value of $\tilde{\rho}(t)$ is given by $\tilde{\rho}(0)=\hat{I}_{{\rm PN}}\otimes\rho(0)$,
where $\hat{I}_{{\rm PN}}$ is the two-dimensional identity matrix and $\rho(0)=\left|gg\right\rangle \left\langle gg\right|$ denotes the initial state of the atom pair. 

The quantum pumping from a single photon pulse is characterized by,
\begin{equation}
\hat{\hat{\mathcal{L}}}_{{\rm pump}}(t)\tilde{\rho}(t) =  \sum_{j=1,2} \sqrt{\gamma_{jj}}\eta_{j}  \left\{ \xi^{*}(t  -  t_{j}) \left[\hat{\sigma}_{j+},\tilde{\rho}(t)\hat{\tau}_{-}\right]  +  \rm{h.c.}   \right\} ,
\end{equation}
where $\gamma_{jj}=\gamma_{0}$ is the spontaneous decay rate of the atoms in vacuum. The coefficient $\eta_j$ characterizes the pumping efficiency, which is determined by the effective scattering cross section of the $j$th atom. The wave-packet amplitude of a Gaussian single photon pulse is given by
\begin{equation}
\xi(t)=\left(\frac{1}{2\pi\tau_f^2}\right)^{1/4}\exp\left[-\frac{t^2}{4\tau_f^2}-i\omega_0 t \right],    
\end{equation}
with center frequency $\omega_0$ and pulse length $\tau_f$~\cite{yang2018concept}. The time that the center of the pulse arrives at the $j$th atom is given by $t_j=\boldsymbol{k}_0\cdot\boldsymbol{x}_j/\omega_0$ ($|\boldsymbol{k}_0|=\omega_0/c$). The absorption of the pulse is characterized by the Pauli matrix $\hat{\tau}_{-}$ of the extra qubit degree. The interatomic RDDI are included in the regular time-independent Lindblad superoperator~\cite{ficek2002entangled,Dung2002resonant}
\begin{align}
\hat{\hat{\mathcal{L}}}_{{\rm atom}}\tilde{\rho}(t) = & -i\left[\sum_{j=1,2}\omega_{0}\hat{\sigma}_{j}^{+}\hat{\sigma}_{j}^{-}
+\sum_{i,j}\delta_{ij}\hat{\sigma}_{i}^{+}\hat{\sigma}_{j}^{-},\tilde{\rho}(t)\right]\nonumber \\ 
& +\sum_{ij}\frac{1}{2}\gamma_{ij} \left[2\hat{\sigma}_{i}^{-}\tilde{\rho}(t)\hat{\sigma}_{j}^{+}  -\tilde{\rho}(t)\hat{\sigma}_{i}^{+}\hat{\sigma}_{j}^{-}-\hat{\sigma}_{i}^{+}\hat{\sigma}_{j}^{-}\tilde{\rho}(t)\right], \label{Eq:free_evolution}
\end{align}
where $\omega_0$ is the energy splitting of the two-level atoms, and the energy shifts $\delta_{ij} = U_{eg,ge}(r)/
\hbar$ and decay rates $\gamma_{ij}$ are given in ~\ref{sec:MEq}. 

\begin{figure}
\centering
\includegraphics[width=7cm]{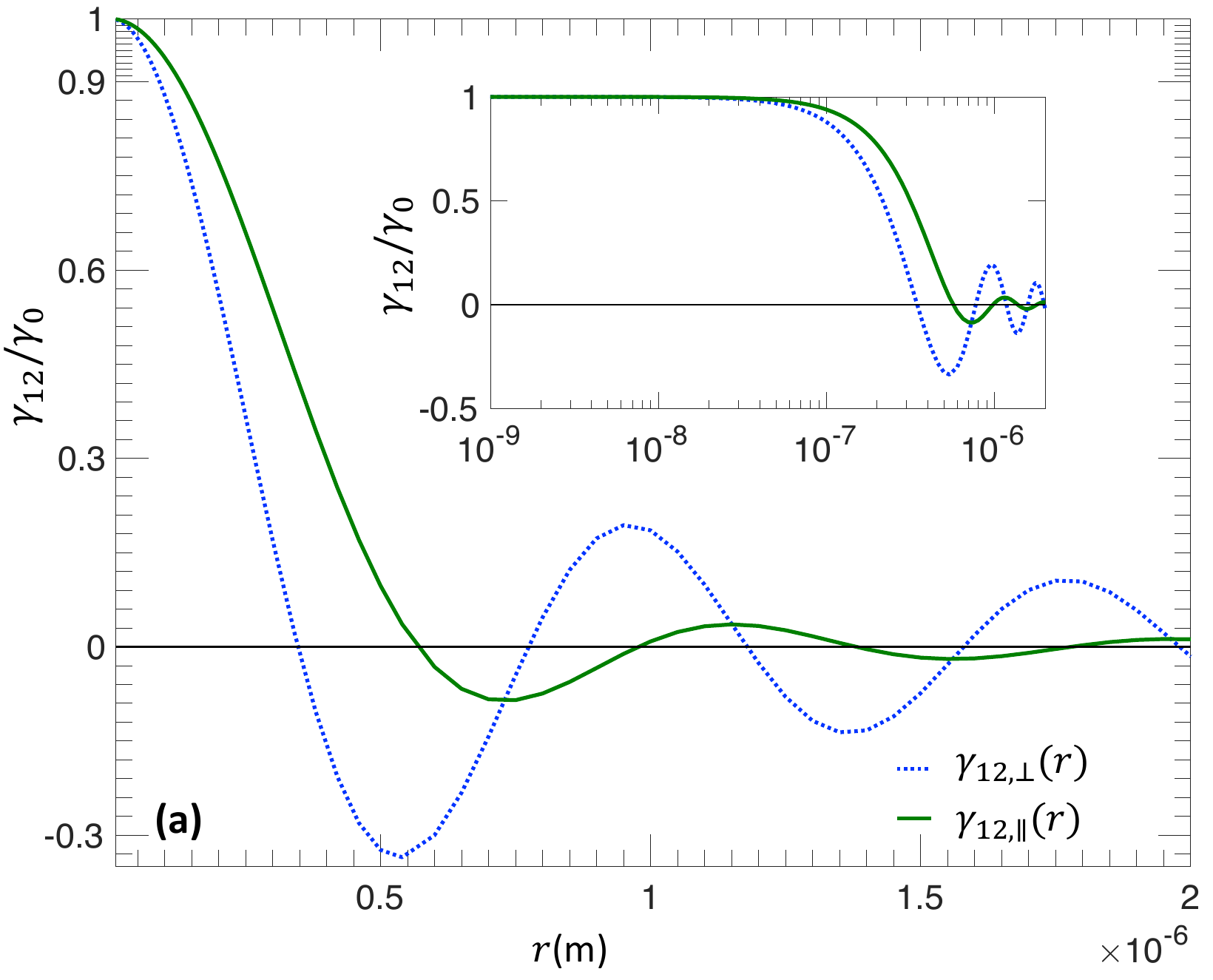}\includegraphics[width=7cm]{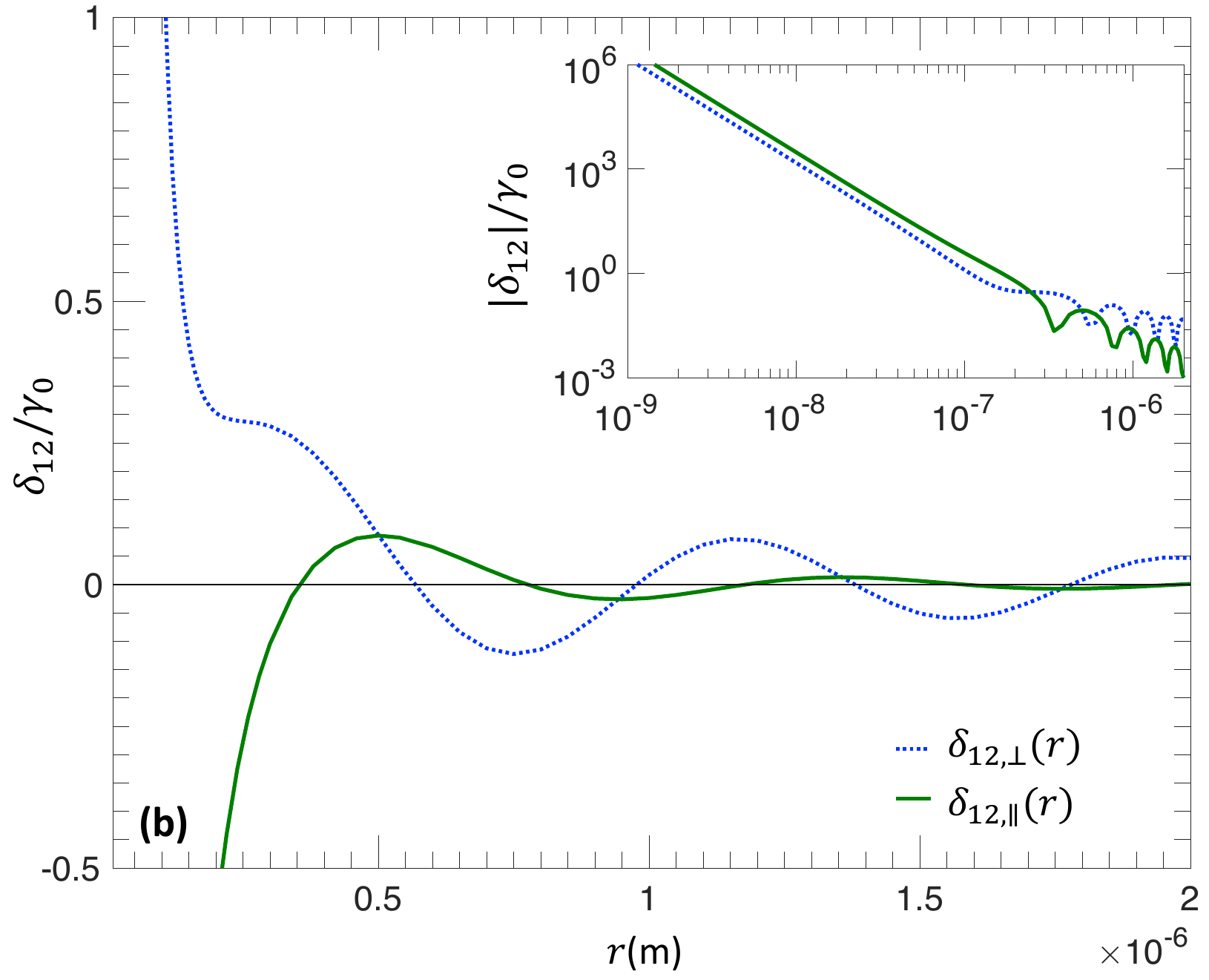}\caption{\label{fig:rddi} Then incoherent part {[}cooperative decay rates
(a){]} and coherent part {[}the energy shifts (b){]} of the RDDI in free space. The sub-indices $\parallel$
and $\perp$ denote the RDDI triggered by parallelly and perpendicularly
polarized (with respect to the atom co-axis) single-photon pulse. In
the subplot, we plot the $r$-axis in log scale. }
\end{figure}

Both the imaginary part (the cooperative decay rates $\gamma_{12}=\gamma_{21}$)  and the real part  (the energy shift $\delta_{12}=\delta_{21}$) of the RDDI are dependent on the polarization of the atomic dipoles $\boldsymbol{\mu}_j$ with respect to the  relative  displacement vector $\boldsymbol{r}$. As shown in Fig.~\ref{fig:rddi}(a), the cooperative decay rates decrease monotonously with atom-atom distance $r$ in the near region, begins to oscillate in the medium region, and vanishes in the far region. Note that, the sub-indices $_{\parallel}$ and $_{\perp}$ denote the cases when $\boldsymbol{\mu}_j$ is parallel and perpendicular to $\boldsymbol{r}$, respectively. Although $\gamma_{12,\parallel}$ and $\gamma_{12,\perp}$ behave differently, both of them converges to the spontaneous decay
rate $\gamma_{0}$ in the near region and decrease to zero in the far region {[}see the subplot in Fig.~\ref{fig:rddi}(a){]}. Rewriting the master equation (\ref{Eq:free_evolution}) in the bright and dark states basis, this will automatically give the superradiance and subradiance~\cite{DeVoe1996observation}.
The coherent part of the RDDI diverges in the near region. More importantly, $\delta_{12,\parallel}$ and $\delta_{12,\perp}$ usually have opposite signs, especially in the near region. This lays the
foundation to tune the RDDI force by tuning the polarization of the pulse as explained in the following.

\begin{figure}
\centering
\includegraphics[width=9cm]{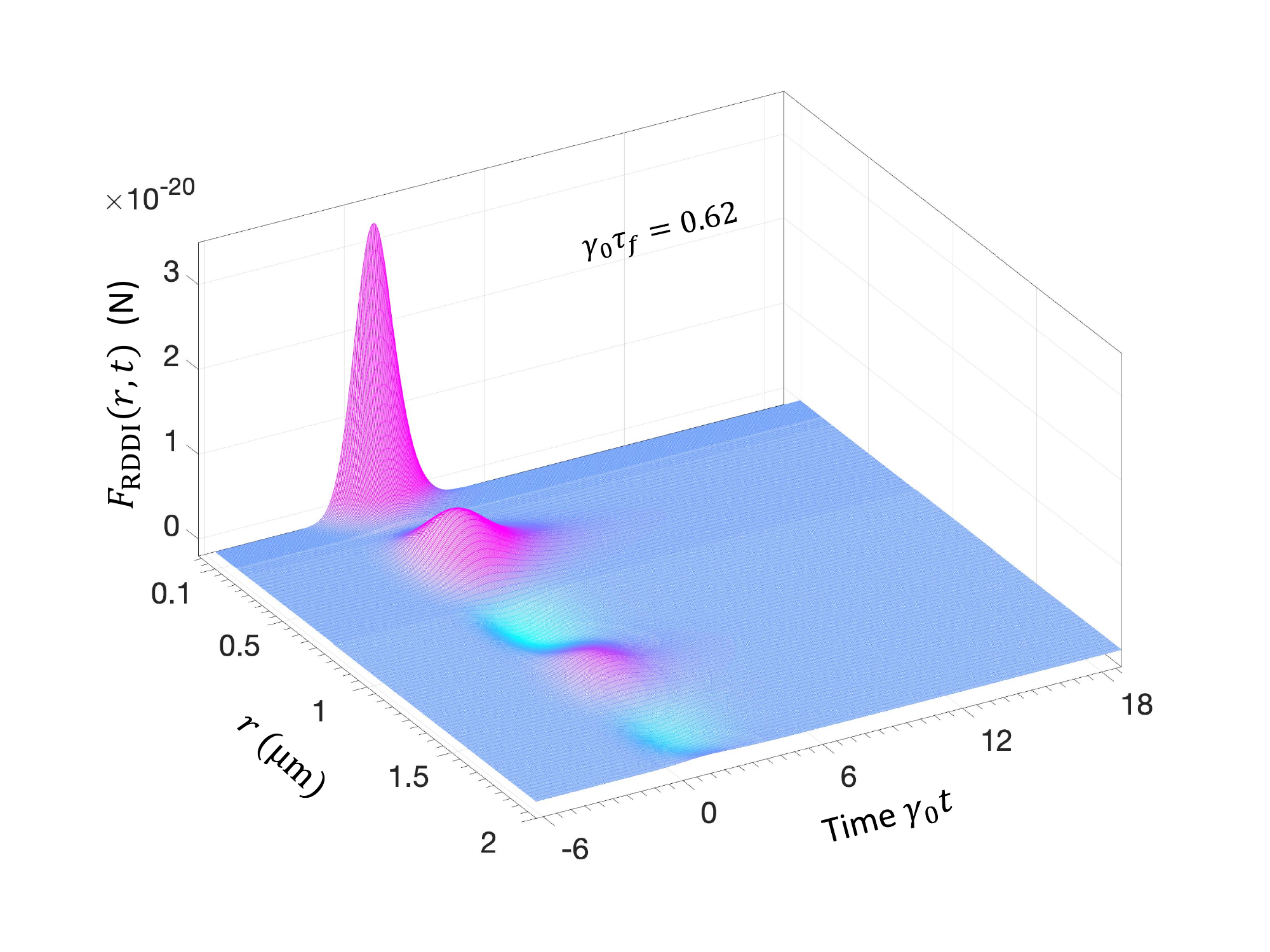}
\caption{\label{fig:2} Single-photon pulse induced transient entanglement force between two ${\rm Rb}$ atoms (D1 transition from $5^{2}S_{1/2}\rightarrow5^{2}P_{1/2}$).  The force reaches its maximum when the photon absorption probability is largest. The magnitude of the RDDI force oscillates with atom-atom distance around $r\sim1\:{\rm \mu m}$. Here, the time is in the unit of $1/\gamma_{0}$ ($\gamma_{0}$ is spontaneous decay rate of the atom in free space). Perpendicularly  polarized pulse ($\perp$) is selected and its pulse length is set as $\gamma_{0}\tau_{f}=0.63$.  The pumping efficiency is set as $\eta_1=\eta_2=1/\sqrt{2}$. The exact data of the ${\rm Rb}$ atom is given in Table~\ref{tab:1}.}
\end{figure}

The time-dependent RDDI entanglement force, $F_{{\rm RDDI}}(r,t)=Tr[\rho(t)\hat{F}_{\rm RDDI}(r)]$, induced by a single photon pulse for different atom distance is displayed in Fig.~\ref{fig:2}. For a fixed inter-atomic distance, the RDDI force increases after the pulse excites the atoms and decreases with time when atoms re-emit the photon. We can also see the amplitude of the RDDI force oscillates with atom distance $r$, due to the oscillation in the matrix elements $F_{ge,eg}(r)$ of the RDDI force operator. The van der Waals force has been neglected here as it is negligibly small as shown in~\ref{sec:vdW}. The impulse force from the incident pulse is estimated to be $F_{\rm imp}\approx \hbar\omega_0/c\tau_f \sim10^{-20}$N with center frequency $\omega_{0}\approx2\pi\times3.77\times10^{14}\:{\rm Hz}$ and pulse length $\tau_{f}\sim30\,{\rm ns}$. But this force is along $y$-axis, which is perpendicular to the inter-atomic
force in $x$-direction and can be relieved by the trapping force in $y$-axis. Thus, the only relevant force along the axis joining the two atoms is the RDDI entanglement force.

\begin{figure}
\centering
\includegraphics[width=12cm]{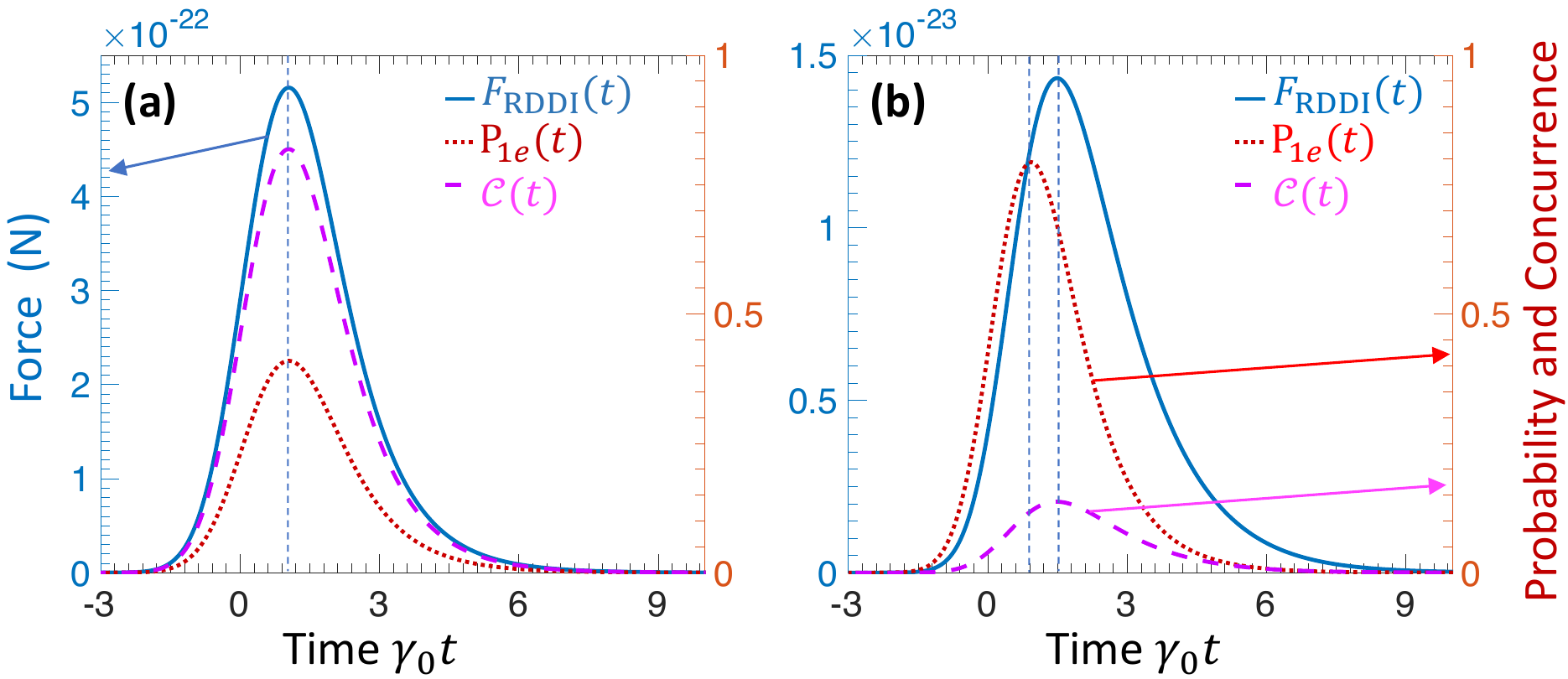}\caption{\label{fig:3}The transient entangled force $F_{{\rm RDDI}}(t)$ (the solid-blue
line), the concurrence $\mathcal{C}(t)$ (the dashed-pink line), and the
excitation probability of the first atom $P_{1e}(t)$ (the dotted-red
line) induced by single-photon pulse.  (a) All the three quantities reach the maximum at the same time in the homogeneous pumping case with pumping efficiency $\eta_1 = \eta_2 = 1/\sqrt{2}$ and pulse length $\tau_{f}\gamma_{0}=0.62$. Thus, the entanglement is generated by the single photon pulse. (b) The excitation probability $P_{1e}(t)$ first reaches its maximum and then the force and the
concurrence reach their maximum in the local pumping case with $\eta_1=1$, $\eta_2 = 0$, and $\tau_{f}\gamma_{0}/2\pi=0.75$. Thus, the two-body entanglement
is generated via the dipole-dipole interaction. Here, the atom-atom distance is fixed as
$r=1.2\,{\rm \mu m}$. In the double-$y$-axis figure, $F_{{\rm RDDI}}(t)$ is associated with the left $y$-axis and both $\mathcal{C}(t)$ and
$P_{1e}(t)$ are associated with the right $y$-axis.}
\end{figure}

Quantum entanglement fundamentally determines the time-dependent RDDI force induced by a single photon pulse. Here, we use the concurrence to quantitatively characterize the two-qubit entanglement~\cite{Wootters1998entanglement}. As shown in Fig.~\ref{fig:3} (a), for fixed atom-atom distance $r=1.2\,{\rm \mu m}$, the concurrence~$\mathcal{C}(t)$(the dashed-pink line) and the RDDI force $F_{{\rm RDDI}}(t)$ (the solid-blue line), as well as the excitation probability of the first atom $P_{1e}(t)$ (the dotted-red line), reach their maxima simultaneously for homogeneous pumping case ($\eta_1=\eta_2$). But for the local pumping of the first atom case with $\eta_1 = 1$ and $\eta_2=0$ [see Fig.~\ref{fig:3} (b)], $\mathcal{C}(t)$ and $F_{{\rm RDDI}}(t)$ reach their peaks at the time, which is later than the time when $P_{1e}(t)$ reaches its maximum. Thus, it is the entanglement instead of the total excitation probability that maximizes the RDDI force. We also see that there are two ways to generate the quantum entanglement between the
atoms: (1) homogeneous pumping  to the symmetric state $\left|\Psi^{+}\right\rangle $ directly by the single photon pulse; (2) local pumping  of single atom to state $\left|eg\right\rangle $ and then the RDDI evolves the atoms to entangled states. Here, we show that the first one is more efficient for entanglement generation. The total photon absorption probability $P_{e,tot}(t)$ for both homogeneous [$P_{e,tot}(t)=2P_{1e}(t)$ in Fig.~\ref{fig:3}(a)] and local pumping cases [$P_{e,tot}(t)=P_{1e}(t)$ in Fig.~\ref{fig:3}(b)] are almost the same. But the entanglement and the RDDI entanglement force under homogeneous pumping are much larger than that of local pumping case. This is because the projection of the atomic state $\rho(t)$ on the entangled state $\left|\Psi^{+}\right\rangle $ under homogeneous pumping is much larger.

\begin{figure}
\centering
\includegraphics[width=12cm]{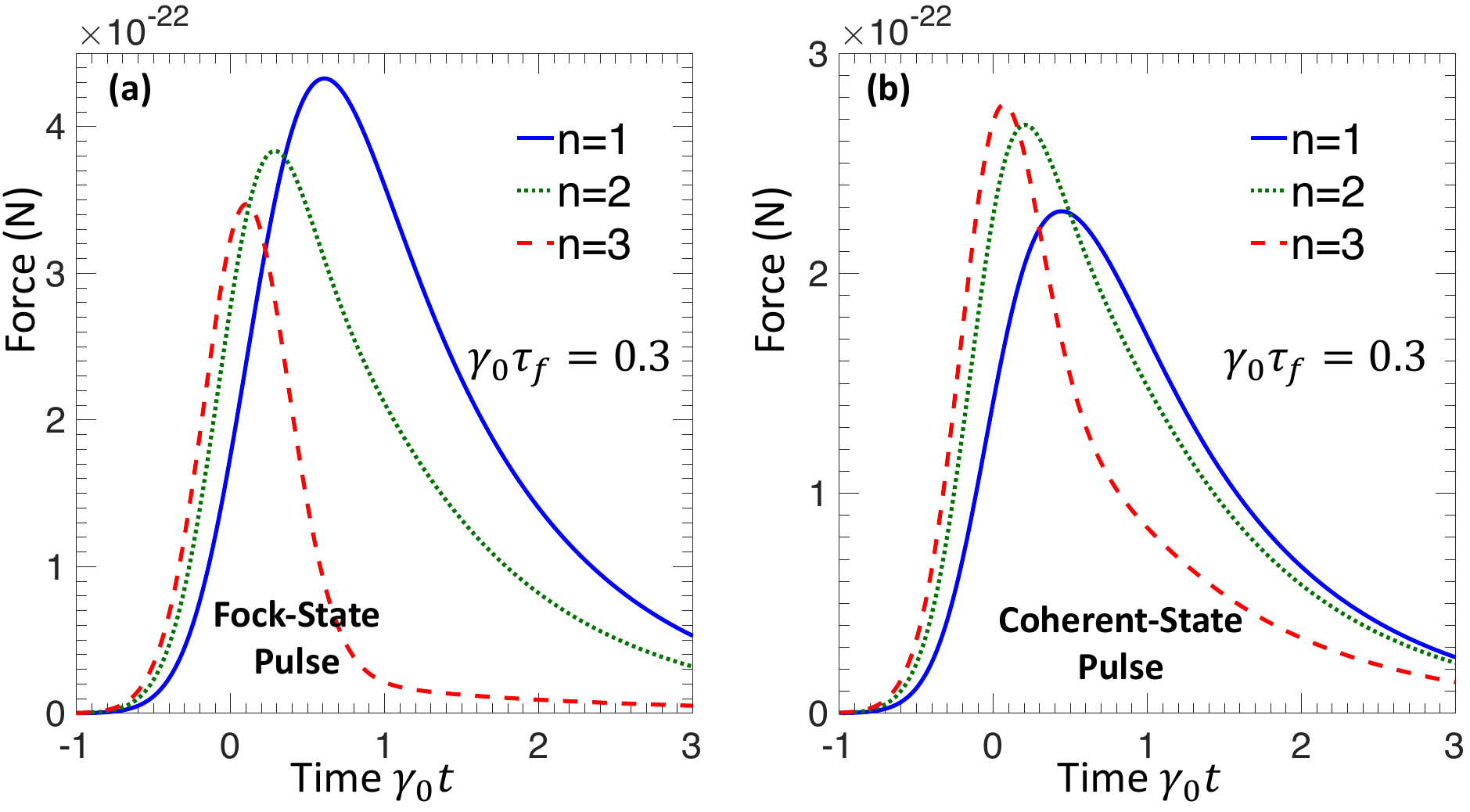}
\caption{\label{fig:4}Comparison of the entanglement force induced by (a) Fock-state pulses and (b) coherent pulses. The 
Fock-state pulse induced force decreases with photon number ($n$) for the fixed pulse length $\tau_{f}\gamma_0 =0.3$, while coherent-state pulses induced force increases with the mean photon number from $1$ to $10$. Here, the
atom-atom distance is fixed at $r=1.2\,{\rm \mu m}$ and $\eta_1=\eta_2 = 1/\sqrt{2}$.}
\end{figure}

The existing theory~\cite{craig1998molecular,buhmann2007dispersion,Dung2002resonant,ficek2002entangled} can not describe the  quantum pulse induced dipole-dipole interaction force. Now, we show that the force induced by a Fock-state pulse is significantly different from the one induced by a coherent-state pulse. As explained in Ref.~\cite{yang2018concept}, the absorption probability of Fock-state single photon pulse by a two-level atom is much higher than that of coherent-state pulse. Thus, the corresponding force is larger as shown by the blue lines in Fig.~\ref{fig:4}. However, there exists an optimal pulse length $\tau_{f,{\rm opt}}$ to reach the largest excitation probability of the atoms for Fock-state pulses~\cite{Baragiola2012n-photon}. For fixed pulse length $\tau_f\gamma_0 = 0.3$, the maximum entanglement force decreases with photon number in Fig.~\ref{fig:4} (a), as the total excitation probability decreases~\cite{Baragiola2012n-photon}. But the force induced by coherent pulse always increases with the mean photon number [see Fig.~\ref{fig:4} (b)]. In an experiment, larger entanglement force can be obtained by optimizing the pulse length to increase the atomic excitation probability for given atomic transition frequency and dipole-dipole interaction strength as shown in~\ref{sec:TDMEQ}.

\section{Near-field enhancement of the entanglement force}

The entanglement force can be enhanced significantly by engineering the nanophotonic environment near the atoms. As a practical illustration, we demonstrate this enhancement by placing the atoms near a graphene layer as depicted in Fig.~\ref{fig:1} (b). The surface plasmon polaritons of graphene have been previously shown to allow conventionally forbidden atomic transitions~\cite{rivera2016shrinking} in addition to enhancing other well-known physical effects such as decay rate of emitters~\cite{koppens2011graphene} and F{\"o}rster energy transfer rate~\cite{biehs2013large}. This enhancement fundamentally originates from the strong light-matter interaction due to the large density of states of the surface plasmon modes, i.e. the polaritons generated by the strong coupling between the electromagnetic field and the charge excitations at a conductor surface~\cite{koppens2011graphene}. Since the field is strongly confined at the surface, thus the corresponding enhancement only occurs when the emitters are placed close to the surface.

Here, we show that the RDDI strength and the time-dependent entanglement force can  be enhanced significantly by placing the atoms near a graphene layer. As presented in~\ref{sec:MEq}, the RDDI strength can be directly evaluated via the classical Green's tensor $\overleftrightarrow{G}(\mathbf{x}_{1},\mathbf{x}_{2},\omega)$. In the presence of a planar surface, the Green tensor in the upper half-space can always be split into two parts~\cite{novotny2012principles}: $\overleftrightarrow{G}(\mathbf{x}_{1},\mathbf{x}_{2},\omega)=\overleftrightarrow{G}_0(\mathbf{x}_{1},\mathbf{x}_{2},\omega)+\overleftrightarrow{G}_R(\mathbf{x}_{1},\mathbf{x}_{2},\omega)$ corresponding to the contributions from the free space and the reflection by graphene, respectively. The free space Green tensor has been analytically given in Refs.~\cite{knoll2000qed,Safari2015body,Morice1995refractive}. The reflection Green tensor can be obtained from the optical conductivity of a graphene layer (see more details in~\ref{sec:graphene}). The in-plane optical conductivity of graphene includes intra-band and inter-band contributions~\cite{koppens2011graphene,biehs2013large,nikitin2011fields,falkovsky2007space,stauber2008optical} $\sigma(\omega)=\sigma_{{\rm intra}}(\omega)+\sigma_{{\rm inter}}(\omega)$ with
\begin{align}
\sigma_{{\rm intra}}(\omega) & =\frac{2e^{2}k_{B}T}{\pi\hbar^{2}}\frac{i}{\omega+i/\tau_{D}}\log\left[2\cosh(E_{F}/2k_{B}T)\right],
\end{align}
and
\begin{align}
\sigma_{{\rm inter}}(\omega) & =\frac{e^{2}}{4\hbar}\left[H(\hbar\omega/2)+\frac{4i\hbar\omega}{\pi}\int_{0}^{\infty}dx\frac{H(x)-H(\hbar\omega/2)}{\hbar^{2}\omega^{2}-4x^{2}}\right],
\end{align}
where $\tau_D$ is the relaxation time in the Drude model, $E_{F}$ is the graphene's Fermi energy, $T$ is the temperature, and the function
\begin{align}
H(x) & =\frac{\sinh(x/k_{B}T)}{\cosh(E_{F}/k_{B}T)+\cosh(x/k_{B}T)}.
\end{align} 

\begin{figure}
\centering
\includegraphics[width=12cm]{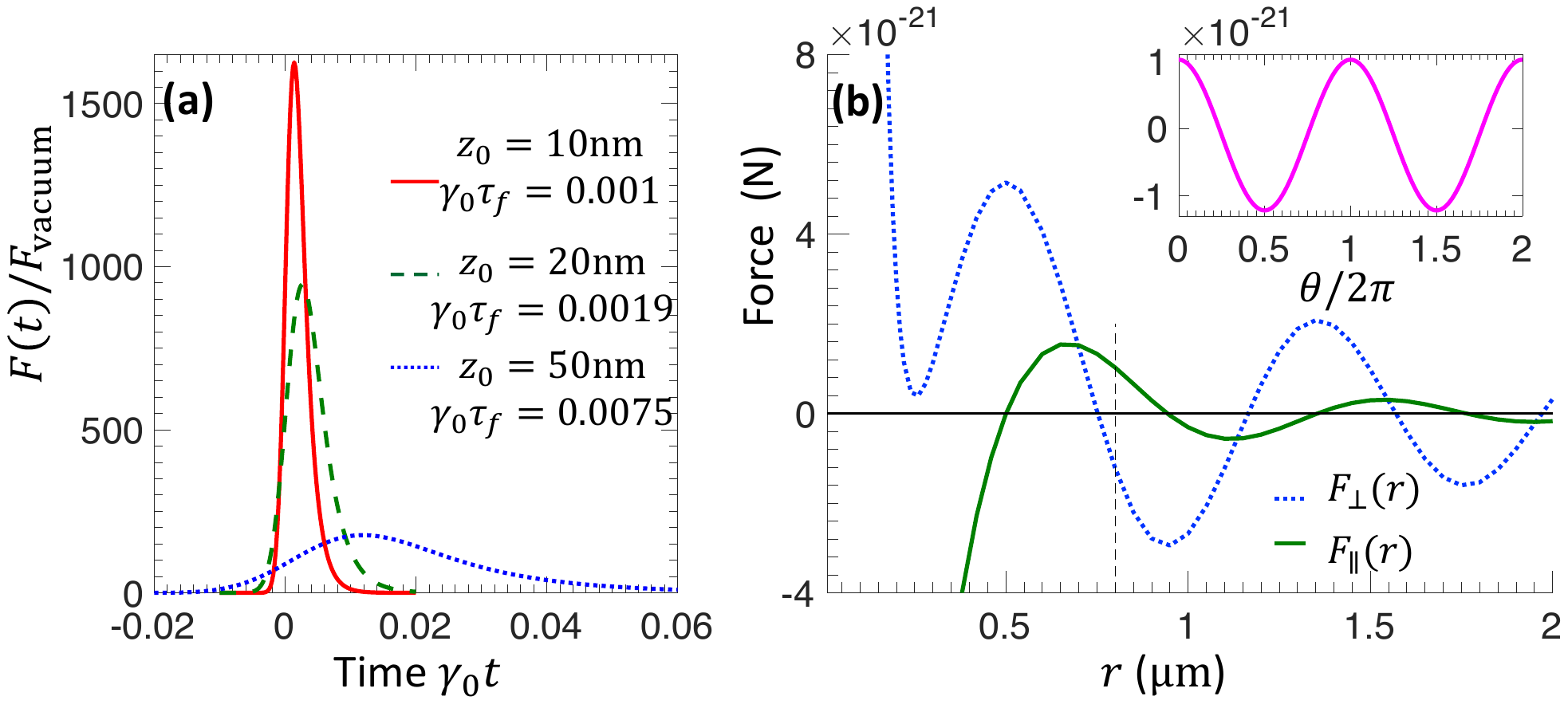}\caption{\label{fig:5} (a) Single-photon pulse induced entanglement force between two atoms placed near a graphene-layer interface. Here, the forces have been normalized by the eigenvalue of the force operator $\hat{F}_{\rm RDDI}(r)$ in free space.  (b) Eigen value of the force operator $\hat{F}_{\rm RDDI}(r)$ for two ${\rm Rb}$ atoms in free space as a function of atomic distance $r$. The induced RDDI forces $F_{{\rm RDDI}}$ are different for parallel ($\parallel$) and perpendicular ($\perp$) polarizations of single-photon pulses, as shown by the solid-green ($F_{\parallel}$) and dotted-blue ($F_{\perp}$) curves. In the subplot, we show the force $F_{{\rm RDDI}}$ with $r=0.8\,{\rm \mu m}$ (marked by the vertical dashed line) for different polarization angle
($\theta$ with respect to $x$-axis) of the pulse in $xz$-plane. This clearly shows the change in sign of the force from repulsive to attractive.}
\end{figure}

Figure~\ref{fig:5} (a) demonstrates the distance dependence of the entanglement force. For atomic transition frequency close to graphene surface plasmon polaritons (exact data provided in~\ref{sec:graphene}), the enhancement factor is larger than $1000$ at atom-surface distance $z_0=10$~nm (red curve). When the two atoms are very close to the graphene layer, the RDDI is primarily mediated by the surface plasmon polaritons in the graphene layer instead of the vacuum fluctuations. The large density of states of surface polaritons enhances the strength of RDDI by orders of magnitude. While the graphene-based surface plasmon polaritons occur in the terahertz to near-infrared band~\cite{koppens2011graphene,nikitin2011fields}, similar enhancement at optical frequencies are feasible with other plasmonic materials such as gold and silver~\cite{west2010searching}. 

\section{Precise control of the entanglement force}

Now, we show single photon pulse as a novel tool to precisely control the atomic force: (1) a more than ten orders of dipole-dipole interaction force amplitude change can be induced by a single photon pulse; (2) the induced entanglement force can be continuously tuned from being repulsive to attractive by varying the polarization of the pulse. For relevant inter-atomic separations ($r\sim1\,{\rm \mu m}$), the van der Waals force is around $\sim5\times10^{-35}$~N (see Fig.~\ref{fig:S3}), which is far beyond the state-of-art force sensitivity. As the van der Waals force arises from higher-order process, thus it is much weaker than the RDDI force. After absorption of a single photon pulse, the RDDI force dominates with a greatly enhanced amplitude $\sim 10^{-22}$~N. This force can be further enhanced upto $10^{-19}$~N with surface plasmons-plaritons. Using phase-coherent Doppler velocimetry, force sensitivity of $\sim10^{-24}{\rm N/\sqrt{Hz}}$ can be approached in trapped ion systems~\cite{biercuk2010ultrasensitive}. In a Mach-Zehnder-type interferometer with a free fall cesium atom from an optical
tweezer, a force of magnitude $3.2\times 10^{-27}$~N has been measured in an experiment~\cite{Parazzoli2012observation}. Therefore, we are confident that that the transient entanglement force induced by a single-photon pulse can also be detected in the near future.

For atomic transition between states connected by linearly polarized light, the direction of the corresponding transition dipole is determined by the polarization of the incident pulse. As shown in Fig.~\ref{fig:5} (b), both the forces induced by parallelly ($\parallel$) and perpendicularly ($\perp$) polarized pulses  oscillate with the atomic distance around $r\sim1\,{\rm \mu m}$. But these two forces have a phase shift and usually have opposite signs (especially in the near region $r<0.5\,{\rm \mu m}$). Thus, we can control the force to be repulsive or attractive by changing only the polarization of the pulse. More importantly, we can continuously tune the value of the RDDI force via tuning the pulse polarization angle $\theta$ in $xz$-plane with fixed atom-atom distance ($r$) (see the subplot).

\section*{Conclusion and discussion}
We reveals the essential role of the two-body entanglement in the RDDI force. We utilize a time-dependent theoretical framework to study the transient entanglement force between two neutral atoms induced by a quantum pulse. We also show that this entanglement force can be significantly enhanced by engineering their nano-photonic environment and precisely controlled by tuning the polarization of the incident pulse.

Looking ahead, our work provides a natural platform to investigate photoassociation in chemical reactions and bioprocesses ~\cite{liu2018building}. By generalizing the force operator to multi-atom case, we can also study the role of the many-body entanglement in the collective force of neutral atom ensemble\cite{Sinha2018collective,yang2018concept}. The photon absorption probability and atom-atom entanglement can be enhanced by tailoring the shape and the time-frequency correlation of photon pulses~\cite{yang2018concept}.      

\section*{Acknowledgments}
This work is supported by the DARPA DETECT ARO award (W911NF-18-1-0074).

\appendix
\section{Dipole-dipole interaction force operator\label{sec:force_def}}
According to the Hellmann-Feynman theorem~\cite{Feynman1937force}, we perform
the derivation to the secular equation with respect to the atom-atom separation $r$,
\begin{equation}
\hat{H}\left|n\right\rangle =\left(\sum_{l}H_{lk}\left|l\right\rangle \left\langle k\right|\right)\left|n\right\rangle =\sum_{l}H_{ln}\left|l\right\rangle 
\end{equation}
 to obtain
\begin{equation}
\left(\frac{\partial}{\partial r}\hat{H}\right)\left|n\right\rangle +\hat{H}\left|\frac{\partial}{\partial r}n\right\rangle =\sum_{l}\left[\left(\frac{\partial}{\partial r}H_{ln}\right)\left|l\right\rangle +H_{ln}\left|\frac{\partial}{\partial r}l\right\rangle \right].
\end{equation}
Multiply both side with $\left\langle m\right|$, we have 
\begin{equation}
\left\langle m\right|\left(\frac{\partial}{\partial r}\hat{H}\right)\left|n\right\rangle =\frac{\partial}{\partial r}H_{mn}+\sum_{l}\left[H_{ln}\left\langle m\right|\frac{\partial}{\partial r}l\rangle-H_{ml}\left\langle l\right|\frac{\partial}{\partial r}n\rangle\right].
\end{equation}
In most case, due to the non-adiabatic transition terms in the square
brackets, there does not exist a well defined force operator for a microscopic system, such as the exchanging interaction in a condensed-matter lattice. But in our case, the distance between the two atoms is much larger than the size the the atoms. Thus, the atomic wave function is not dependent on the relative distance $r$ and the second term  at the right-hand-side disappears (i.e., $\langle l|\partial n/\partial r\rangle=0$). 

In the atomic Hamiltonian, only the dipole-dipole interaction part 
\begin{equation}
\hat{U}(r)=U_{mn}(r)|m\rangle\langle n|,\label{eq:Ur}    
\end{equation}
depends on the inter-atomic distance $r$. As the corresponding force is always along the co-axis line, we can define a scalar operator for this force as,
\begin{align}
\hat{F}(r) & \equiv -\frac{\partial}{\partial r}\hat{H}= -\sum_{mn}\left[\frac{\partial}{\partial r}U_{mn}(r)\right]\left|m\right\rangle \left\langle n\right|.
\end{align}

We note that this force operator only works for weak atom-field coupling case. If the two atoms strongly coupled to a resonant cavity field, one can not eliminate the degree of the cavity mode to obtain an effective interaction Hamiltonian as shown in Eq.~(\ref{eq:Ur}). In this case, the inter-atomic force is not only dependent on atom-atom separation, but also the position of each atom~\cite{esfandiarpour2018cavity}. More important, the magnitude of the forces experienced by the two atoms can be different, which violates Newton's third law for a macroscopic body. We do not consider this case in this paper.
 
Different elements in the operator $\hat{F}(r)$ correspond to different virtual processes generated forces. We emphasize that only the anti-diagonal elements of the two-body interaction in (\ref{eq:Ur}) can be mediated by second-order processes~\cite{craig1998molecular} and all the other terms result mainly from fourth order processes. Thus, the corresponding forces are weak. In this paper, we only focus on two forces. The first one is the van der Waals (vdW) force between two ground-state atoms $F_{{\rm vdW}}\propto F_{gg,gg}(r)$, which mainly arises from fourth-order
process~\cite{craig1998molecular,Safari2006body} and usually is extremely small. An incident single-photon pulse can pump the atom pair to an entangled state. In this case, the interaction changes to the RDDI, which plays the key role in energy transfer between different molecules in chemical and biological processes. As the RDDI is mediated by second-order processes, the corresponding force $F_{{\rm RDDI}}\sim F_{ge,eg}(r)$ between the two atoms will be greatly enhanced.  In the following, we present the approach to calculate the elements of $\hat{U}(r)$ and $\hat{F}(r)$.

\section{Model Hamiltonian for Atom-Field Interaction}

The Hamiltonian of the total system is given by 
\begin{equation}
\hat{H}=\hat{H}_{F}+\sum_{j=1,2}\hat{H}_{A,j}+\sum_{j}\hat{H}_{AF,j},
\end{equation}
where the Hamiltonian of the field modes in an arbitrary linear (non-magnetic) media is given by~\cite{Dung1998three,knoll2000qed}
\begin{equation}
\hat{H}_{F}=\int d^{3}\mathbf{x}\int_{0}^{\infty}d\omega\hbar\omega\hat{\mathbf{f}}^{\dagger}(\mathbf{x},\omega)\cdot\hat{\mathbf{f}}(\mathbf{x},\omega),
\end{equation}
and the ladder operators of the eigen modes satisfy the commutation
relations
\begin{equation}
[\hat{f}_{\alpha}(\mathbf{x},\omega),\hat{f}_{\beta}^{\dagger}(\mathbf{x}',\omega')]=\delta_{\alpha\beta}\delta(\mathbf{x}-\mathbf{x}')\delta(\omega-\omega'),\ \alpha,\beta=x,y,z
\end{equation}
and
\begin{equation}
[\hat{f}_{\alpha}(\mathbf{x},\omega),\hat{f}_{\beta}(\mathbf{x}',\omega')]=[\hat{f}^{\dagger}_{\alpha}(\mathbf{x},\omega),\hat{f}_{\beta}^{\dagger}(\mathbf{x}',\omega')]=0.
\end{equation}
The Hamiltonian of the two atoms is 
\begin{equation}
H_{A,j}=\hbar\omega_{a,j}\hat{\sigma}_{j}^{+}\hat{\sigma}_{j}^{-},
\end{equation}
where $\omega_{a,j}$ is the energy splitting of the $j$th atom and
$\hat{\sigma}_{j}^{+}=(\hat{\sigma}_{j}^{-})^{\dagger}=\left|e_{j}\right\rangle \left\langle g_{j}\right|$
is the Puali matrix. There are two forms of Hamiltonian to describe
the interaction between the atoms and the electromagnetic field. One
is the minimum coupling and the other one is the multipolar coupling~\cite{craig1998molecular}.
The difference and relation between these two forms of interaction
can be found in~\cite{craig1998molecular,Buhmann2004Casimir}. Here,
we use the multiploar interaction Hamiltonian
\begin{equation}
\hat{H}_{AF,j}=-(\boldsymbol{\mu}_{j,eg}\hat{\sigma}_{j}^{+}+\boldsymbol{\mu}_{j,ge}\hat{\sigma}_{j}^{-})\cdot\hat{\mathbf{E}}(\mathbf{x}_{j}),\label{eq:Hint}
\end{equation}
where $\boldsymbol{\mu}_{j,eg}$ is the electric dipole transition element
of the $j$th atom. In the following, for simplicity,
we consider two identical atom case $\boldsymbol{\mu}_{j,eg}=\boldsymbol{\mu}_{j,eg}=\boldsymbol{\mu}_j=d_{0}\boldsymbol{e}_{j}$. 

The electric field operator can be expanded with the eigen modes of
the field as
\begin{equation}
\hat{\mathbf{E}}(\mathbf{x})=\int_{0}^{\infty}d\omega\left[\hat{\mathbf{E}}(\mathbf{x},\omega)+\hat{\mathbf{E}}^{\dagger}(\mathbf{x},\omega)\right],
\end{equation}
where
\begin{align}
\hat{\mathbf{E}}(\mathbf{x},\omega) & =i\sqrt{\frac{\hbar}{\pi\varepsilon_{0}}}\frac{\omega^{2}}{c^{2}}\int d^{3}x'\sqrt{\varepsilon_{I}(\boldsymbol{x}',\omega)}\overleftrightarrow{G}(\mathbf{x},\mathbf{x}',\omega)\cdot\hat{\mathbf{f}}(\mathbf{x}',\omega),\\
\hat{\mathbf{E}}^{\dagger}(\mathbf{x},\omega) & =-i\sqrt{\frac{\hbar}{\pi\varepsilon_{0}}}\frac{\omega^{2}}{c^{2}}\int d^{3}x'\sqrt{\varepsilon_{I}(\boldsymbol{x}',\omega)}\hat{\mathbf{f}}^{\dagger}(\mathbf{x}',\omega)\cdot\overleftrightarrow{G}^{\dagger}(\mathbf{x},\mathbf{x}',\omega),\\
\overleftrightarrow{G}^{\dagger}(\mathbf{x},\mathbf{x}',\omega) & =\overleftrightarrow{G}(\mathbf{x}',\mathbf{x},-\omega^{*}),
\end{align}
with $\sqrt{\varepsilon_{I}(\boldsymbol{x}',\omega)}$ the imaginary
part of the complex permittivity $\varepsilon(\boldsymbol{x},\omega)$,
the vacuum permittivity $\varepsilon_{0}$, and the speed
of light $c$ in vacuum. The function $\overleftrightarrow{G}(\mathbf{x},\mathbf{x}',\omega)$
is the classical Green tensor obeying the equation
\begin{equation}
\left[\vec{\nabla}\times\vec{\nabla}\times-\frac{\omega^{2}}{c^{2}}\varepsilon(\mathbf{x},\omega)\right]\overleftrightarrow{G}(\mathbf{x},\mathbf{x}',\omega)=\overleftrightarrow{I}\delta(\mathbf{x}-\mathbf{x}').
\end{equation}
Here, we assume that the media is a non-magnetic material with constant
permeability $\mu_{0}=1$ and the frequency-dependent complex permittivity
$\varepsilon(\mathbf{x},\omega)$. The Green tensor has the properties
\begin{align}
\overleftrightarrow{G}^{*}(\mathbf{x},\mathbf{x}',\omega) & =\overleftrightarrow{G}(\mathbf{x},\mathbf{x}',-\omega^{*}),\\
\overleftrightarrow{G}^{T}(\mathbf{x},\mathbf{x}',\omega) & =\overleftrightarrow{G}(\mathbf{x}',\mathbf{x},\omega),\\
\int d^{3}x \frac{\omega^{2}}{c^{2}}\varepsilon_{I}(\boldsymbol{x},\omega)\overleftrightarrow{G}(\mathbf{x}_{1},\mathbf{x},\omega)\overleftrightarrow{G}^{\dagger}(\mathbf{x}_{2},\mathbf{x},\omega) & = {\rm Im}\overleftrightarrow{G}(\mathbf{x}_{1},\mathbf{x}_{2},\omega).\label{eq:propgating_eq}
\end{align}

We will show that both the van der Waals interaction and the resonant dipole-dipole interaction can be easily obtained with the Green tensor.

\section{van der Waals Interaction\label{sec:vdW}}

The van der Waals interaction between two atoms has between well studied. A detailed calculation of the coherent van der Waals interaction in free space is presented in Ref.~\cite{craig1998molecular}. Here, we only present
the more general form of van der Waals interaction between two identical
atoms obtained by Safari and his collaborators~\cite{Safari2006body},

\begin{equation}
U_{gg,gg}(r)=-\frac{2\mu_{0}^{2}}{\hbar\pi}\int_{0}^{\infty}du\frac{\omega_{a,1}\omega_{a,2}u^{4}}{[\omega_{a,1}^{2}+u^{2}][\omega_{a,2}^{2}+u^{2}]}[\boldsymbol{\mu}_{1}\cdot\overleftrightarrow{G}(\mathbf{x}_{1},\mathbf{x}_{2},iu)\cdot\boldsymbol{\mu}_{2}]^{2}.
\end{equation}

The incoherent part of van der Waals interaction has been neglected, as it is usually negligible small compared to the spontaneous decay rate of the atoms.

\subsection{Free-space case}
In this subsection, we recover the well known van der Waals force in free space. It is easy to find that if we
let $\mathbf{r}=\mathbf{x}_{2}-\mathbf{x}_{1}=(r,0,0)$, only
the diagonal elements of the free space Green tensor are nonzero~\cite{knoll2000qed,Safari2015body},
\begin{align}
G_{\parallel}(\mathbf{x}_{2},\mathbf{x}_{1},\omega) & =\frac{c^{2}}{2\pi\omega^{2}r^{3}}(1-i\frac{\omega r}{c})e^{i\omega r/c},\label{eq:GTp}\\
G_{\perp}(\mathbf{x}_{2},\mathbf{x}_{1},\omega) & =-\frac{c^{2}}{4\pi\omega^{2}r^{3}}\left[1-i\omega r/c-\frac{\omega^{2}r^{2}}{c^{2}}\right]e^{i\omega r/c}.\label{eq:GTv}
\end{align}
Here, the sub-indices $\parallel$ and $\perp$ denote parallel and perpendicular to $\boldsymbol{r}$, respectively.

As the ground-state atoms can be excited by arbitrarily polarized
virtual photons. Thus, to calculate the van der Waals interaction,
we need average out the polarization angle by taking the spherically
symmetric polarizability tensor {[}see Eq.~(49) in Ref.~\cite{Safari2006body}{]}.
Finally, the van der Waals interaction between two ground-state atom
is given by
\begin{equation}
U_{gg,gg}(r)=-\frac{2\mu_{0}^{2}d_{0}^{4}}{3\hbar\pi}\int_{0}^{\infty}du\frac{\omega_{0}^{2}u^{4}}{(\omega_{0}^{2}+u^{2})^{2}}{\rm Tr}[\overleftrightarrow{G}(\mathbf{x}_{1},\mathbf{x}_{2},iu)\cdot\overleftrightarrow{G}(\mathbf{x}_{2},\mathbf{x}_{1},iu)].
\end{equation}

\begin{figure}
\centering
\includegraphics[width=10cm]{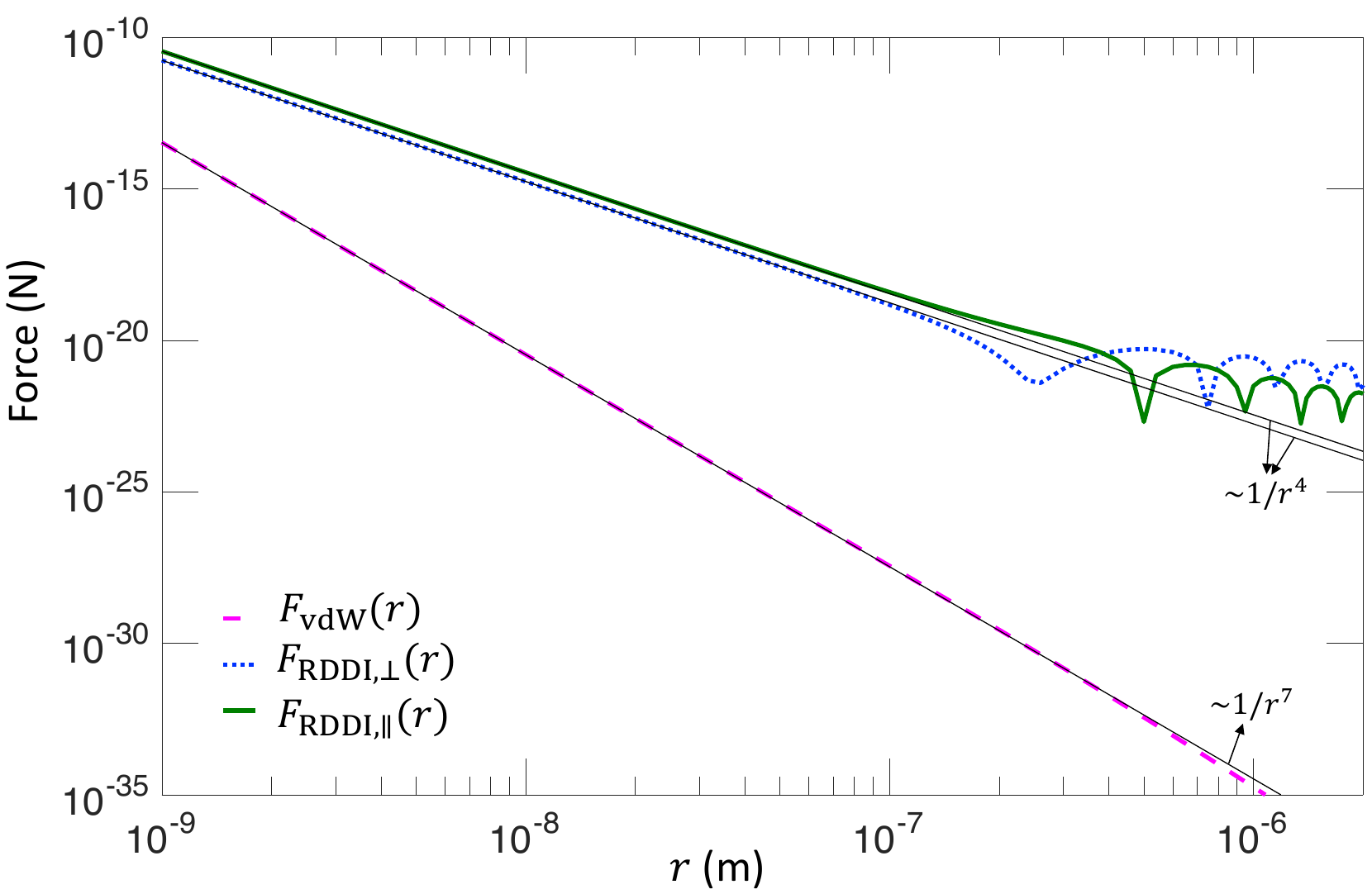}\caption{\label{fig:S3}The matrix element of the dipole-dipole force operator for two Rb atoms. The dashed-pink curve denotes the van der Waals force $F_{\rm vdW}\sim F_{gg,gg}$, which decreases with
the atom-atom distance with scaling $\sim1/r^{7}$ (marked by the thin black line) in the near region and $\sim1/r^{8}$ in the far
region. The eigen value of the RDDI force operator $F_{\rm RDDI}$ is displayed with the dotted-blue line (parallelly polarized atoms $\parallel$) and the solid green line
(perpendicularly polarized atoms $\perp$). The RDDI force decrease with $\sim1/r^{4}$ (marked by
the thin black lines) in the near region and oscillates in the far
region. The data of the two Rb atoms are given in Table~\ref{tab:1}.}
\end{figure}

Using the method presented in~\cite{craig1998molecular} (see Chaps.
7.5 and 7.6), we can verify that:
\begin{equation}
U_{gg,gg}(r)\sim\begin{cases}
1/r^{6}, & ur\ll1\\
1/r^{7}, & ur\gg1
\end{cases}.
\end{equation}
Thus, the corresponding force $F_{{\rm vdW}}(r)$ will be of scale$\sim1/r^{7}$
in the near region and $\sim1/r^{8}$ in the far region. As shown
by the pink curve in Fig.~\ref{fig:S3}, the van der Waals force
$F_{{\rm vdW}}(r)$ deviate from the line $1/r^{7}$ (the thin black
line) slightly in the far region. 

\section{Master-Equation Method to Calculate The Resonant Dipole-Dipole Interaction\label{sec:MEq}}

In this section, we calculate the RDDI strength via the Lindblad form master equation
for a two-level-atom pair
\begin{align}
\frac{d}{dt}\rho(t) & = -i\left[\sum_{j}\omega_{0}\hat{\sigma}_{j}^{+}\hat{\sigma}_{j}^{-} 
+\sum_{i,j}\delta_{ij}\hat{\sigma}_{i}^{+}\hat{\sigma}_{j}^{-},\rho(t)\right]\nonumber\\
& +\sum_{ij}\frac{1}{2}\gamma_{ij}[2\hat{\sigma}_{i}^{-}\rho(t)\hat{\sigma}_{j}^{+}-\rho(t)\hat{\sigma}_{i}^{+}\hat{\sigma}_{j}^{-}-\hat{\sigma}_{i}^{+}\hat{\sigma}_{j}^{-}\rho(t)],\label{eq:MasterEq}
\end{align}
where the decay rates are given by
\begin{align}
\gamma_{ij} & =\frac{2\mu_{0}\omega_{0}^{2}}{\hbar}\boldsymbol{\mu}_i\cdot{\rm Im}\overleftrightarrow{G}(\mathbf{x}_{i},\mathbf{x}_{j},\omega_{0})\cdot\boldsymbol{\mu}_j=\frac{2\omega_{0}^{2}}{\hbar\varepsilon_{0}c^{2}}\boldsymbol{\mu}_i\cdot{\rm Im}\overleftrightarrow{G}(\mathbf{x}_{i},\mathbf{x}_{j},\omega_{0})\cdot\boldsymbol{\mu}_j,\label{eq:gamma_ij}
\end{align}
and the RDDI energy
\begin{align}
\delta_{ij} & =\mathcal{P}\frac{\hbar\mu_{0}}{\pi}\int_{0}^{\infty}d\omega\omega^{2}\times\left[\frac{\boldsymbol{\mu}_i\cdot{\rm Im}\overleftrightarrow{G}(\mathbf{x}_{i},\mathbf{x}_{j},\omega)\cdot\boldsymbol{\mu}_j}{\omega_{0}-\omega}-\frac{\boldsymbol{\mu}_i\cdot{\rm Im}\overleftrightarrow{G}(\mathbf{x}_{j},\mathbf{x}_{i},\omega)\cdot\boldsymbol{\mu}_j}{\omega+\omega_{0}}\right]\label{eq:delta_ij}\\
 & =-\frac{\omega_{0}^{2}}{\hbar\varepsilon_{0}c^{2}}\boldsymbol{\mu}_i\cdot{\rm Re}\overleftrightarrow{G}(\mathbf{x}_{i},\mathbf{x}_{j},\omega_{0})\cdot\boldsymbol{\mu}_j=U_{eg,ge}(r)/\hbar,\label{eq:delta_ij_final}
\end{align}
can also be obtained with Heisenberg equations~\cite{shengwen2018magnetic,supplementary}.

This master equation can also be found
in Refs.~\cite{ficek2002entangled,Dung2002resonant,Dzsotjan2010quantum}. For atomic states connected by linearly polarized light, the direction of the transition dipoles $\boldsymbol{e}_j$ are determined by the polarization of the incident pulse. This makes it possible to precisely control the RDDI force by tuning the polarization of the pulse as shown in the main text.

\begin{table}
\begin{tabular}{|c|c|c|}
\hline 
$^{85}{\rm Rb}$ & Transition frequency $\omega_{0}$ & Wave length\tabularnewline
\hline 
D1 ($5^{2}S_{1/2}\rightarrow5^{2}P_{1/2}$) & $2\pi\times3.77\times10^{14}$ Hz & $794.98$ nm\tabularnewline
\hline 
\hline 
Transition dipole element $d_{0}$ & Spontaneous decay rate $\gamma_{0}$ & Life time $\tau_{0}=1/\gamma_{0}$\tabularnewline
\hline 
$2.54\times10^{-29}\ {\rm C\cdot m}$ & $2\pi\times5.75\times10^{6}$ Hz & $27.68\times10^{-9}\ {\rm s}$\tabularnewline
\hline 
\end{tabular}
\caption{\label{tab:1} The data of the $\rm{^{85}Rb}$ atom used in this paper coming from Ref.~\cite{Ru85data}. We note that the spontaneous decay rate can be obtained directily from Eq.~(\ref{eq:gamma0}) with $\omega_0$ and $d_0$. }
\end{table}

\subsection{Resonant dipole-dipole interaction force in free space}
In this subsection, we calculate the RDDI force in free space. It is straightforward to verify that, for the free space single point Green's
function, the real part diverges, but the imaginary part does not,
\begin{align}
{\rm Im}G_{\parallel}(\mathbf{x}_{1},\mathbf{x}_{1},\omega) & \!=\!\lim_{r\rightarrow0}{\rm Im}\!\!\left[\!\frac{c^{2}}{2\pi\omega^{2}r^{3}}(1\!-\!i\frac{\omega r}{c})e^{i\omega r/c}\!\right]\!=\!\frac{\omega}{6\pi c}\!,
\end{align}
\begin{align}
{\rm Im}G_{\perp}(\mathbf{x}_{1},\mathbf{x}_{1},\omega) & =\frac{\omega}{6\pi c}.
\end{align}
Then, we can obtain the well known spontaneous decay rate of an atoms
in free space,
\begin{align}
\gamma_{11}=\gamma_{22} & =\frac{2\omega_{0}^{2}}{\hbar\varepsilon_{0}c^{2}}\boldsymbol{\mu}_i\cdot{\rm Im}\overleftrightarrow{G}(\mathbf{x}_{i},\mathbf{x}_{j},\omega_{0})\cdot\boldsymbol{\mu}_i=\frac{\omega_{0}^{3}d_{0}^{2}}{3\pi\hbar\varepsilon_{0}c^{3}}\equiv\gamma_{0}. \label{eq:gamma0}
\end{align}
We will take $\gamma_{0}=1$ as the unit of frequency and $1/\gamma_{0}$
as the unit of time in this paper. As shown in the next section, both the coherent and incoherent dipole-dipole interaction can be greatly enhanced by engineering the electromagnetic environment to change the Green tensor.

Substitute the free space Green's tensor (\ref{eq:GTp}) and (\ref{eq:GTv})
back to the incoherent part (\ref{eq:gamma_ij}) and coherent part
(\ref{eq:delta_ij_final}) of the RDDI, we can obtain the corresponding
cooperative decay rates and the energy shifts of the atoms in free space,
\begin{equation}
\gamma_{12,\parallel}=\frac{3}{2}\gamma_{0}\left[-\frac{1}{(k_{0}r)^{3}}\sin(k_{0}r)+\frac{1}{(k_{0}r)^{2}}\cos(k_{0}r)+\frac{1}{k_{0}r}\sin(k_{0}r)\right],
\end{equation}
\begin{equation}
\gamma_{12,\perp}=3\gamma_{0}\left[\frac{1}{(k_{0}r)^{3}}\sin(k_{0}r)-\frac{1}{(k_{0}r)^{2}}\cos(k_{0}r)\right].
\end{equation}
and 
\begin{align}
\delta_{12,\parallel} & =-\frac{3}{2}\hbar\gamma_{0}\left[\frac{1}{(k_{0}r)^{3}}\cos(k_{0}r)+\frac{1}{(k_{0}r)^{2}}\sin(k_{0}r)\right],\\
\delta_{12,\perp} & =\frac{3}{4}\hbar\gamma_{0}\left[\frac{1}{(k_{0}r)^{3}}\cos(k_{0}r)+\frac{1}{(k_{0}r)^{2}}\sin(k_{0}r)-\frac{1}{k_{0}r}\cos(k_{0}r)\right],
\end{align}
where $k_{0}=\omega_{0}/c$. 

The matrix element of the force operator $\hat{F}_{\rm RDDI}$ are given by
\begin{align}
F_{\rm RDDI,\parallel}(r)& = -\frac{\partial}{\partial r}\delta_{12,\parallel}\nonumber \\
& = -\frac{3}{2}\hbar\gamma_{0}\left[\frac{k_{0}}{(k_{0}r)^{4}}\cos(k_{0}r)-\frac{k_{0}}{(k_{0}r)^{3}}\sin(k_{0}r)+\frac{k_{0}}{(k_{0}r)^{2}}\cos(k_{0}r)\right],
\end{align}
and
\begin{align}
F_{\rm RDDI,\perp}(r)& = -\frac{\partial}{\partial r}\delta_{12,\perp}\nonumber \\
& = \frac{3}{4}\hbar\gamma_{0}\left[\frac{k_{0}}{(k_{0}r)^{4}}\cos(k_{0}r)-\frac{2k_{0}}{(k_{0}r)^{2}}\cos(k_{0}r)+\frac{3k_{0}}{(k_{0}r)^{3}}\sin(k_{0}r)-\frac{1}{r}\cos(k_{0}r)\right].
\end{align}
The numerical simulation of the forces are displayed in Fig.~\ref{fig:S3}. In the near region, the RDDI force decreases with $1/r^4$. In the far region, $F_{\rm RDDI,\parallel}$ decreases with $1/r^2$ (green solid line) and $F_{\rm RDDI,\perp}$ vanishes with scaling $1/r$  (blue dotted line).

\section{Dipole-dipole Force near Planar Interface\label{sec:graphene}}
As shown in previous sections, the Green tensor plays the key role in evaluation of the dipole-dipole interaction as well as the corresponding force. In this section, we explain how to calculate the RDDI force near a planar interface via the Green tensor. 

The Green tensor near a planar interface is given by~\cite{novotny2012principles}
\begin{equation}
\overleftrightarrow{G}(\mathbf{x}_{1},\mathbf{x}_{2},\omega)=\begin{cases}
\overleftrightarrow{G}_{0}(\mathbf{x}_{1},\mathbf{x}_{2},\omega)+\overleftrightarrow{G}_{R}(\mathbf{x}_{1},\mathbf{x}_{2},\omega), & z_{1}>0,\ z_{2}>0\\
\overleftrightarrow{G}_{T}(\mathbf{x}_{1},\mathbf{x}_{2},\omega), & z_{1}>0,\ z_{2}<0
\end{cases}
\end{equation}
where $\overleftrightarrow{G}_{0}$ is the Green tensor in the free space, and $\overleftrightarrow{G}_{R}$ and $\overleftrightarrow{G}_{T}$
are the contribution due to the reflection and transmission, respectively. The interface is at the plane $z=0$ and the dipole source (the atoms) are placed above the interface. Thus, all
the reflected field has $z>0$ and all the transmitted field has $z<0$. 

The free-space dyadic Green Tensor in real space can
be written as the sum of the following terms~\cite{arnoldus2003transverse}
\begin{equation}
\overleftrightarrow{G}_{0}(\mathbf{x}_{1},\mathbf{x}_{2},\omega)=\overleftrightarrow{G}_{0}^{{\rm FF}}(\mathbf{x}_{1},\mathbf{x}_{2},\omega)+\overleftrightarrow{G}_{0}^{{\rm IF}}(\mathbf{x}_{1},\mathbf{x}_{2},\omega)+\overleftrightarrow{G}_{0}^{{\rm NF}}(\mathbf{x}_{1},\mathbf{x}_{2},\omega),\label{eq:G0}
\end{equation}
where the far-, intermediate-, and near-field terms are given by,
\begin{equation}
\overleftrightarrow{G}_{0}^{{\rm FF}}(\mathbf{x}_{1},\mathbf{x}_{2},\omega)=\left(\overleftrightarrow{I}-\mathbf{e}_{r}\mathbf{e}_{r}\right)\frac{1}{4\pi r}e^{ik_{\omega}r},
\end{equation}
\begin{equation}
\overleftrightarrow{G}_{0}^{{\rm IF}}(\mathbf{x}_{1},\mathbf{x}_{2},\omega)=i\left(\overleftrightarrow{I}-3\mathbf{e}_{r}\mathbf{e}_{r}\right)\frac{1}{4\pi k_{\omega}r^{2}}e^{ik_{\omega}r},
\end{equation}
and
\begin{equation}
\overleftrightarrow{G}_{0}^{{\rm NF}}(\mathbf{x}_{1},\mathbf{x}_{2},\omega)=-\left(\overleftrightarrow{I}-3\mathbf{e}_{r}\mathbf{e}_{r}\right)\frac{1}{4\pi k_{\omega}^{2}r^{3}}e^{ik_{\omega}r},
\end{equation}
with $\mathbf{e}_{r}=\mathbf{r}/r$. The Green tensor $\overleftrightarrow{G}_{0}$ in (\ref{eq:G0}) is the exact same as the one given in Eqs.~(\ref{eq:GTp}) and~(\ref{eq:GTv}). 

Usually, the reflection Green tenor $\overleftrightarrow{G}_{R} = \overleftrightarrow{G}_{R}^s+\overleftrightarrow{G}_{R}^p$ and the transmission Green tensor $\overleftrightarrow{G}_{T}=\overleftrightarrow{G}_{T}^s+\overleftrightarrow{G}_{T}^p$ (the index $s$ and $p$ denote the s-polarized part and the p-polarized part, respectively) can only be obtained numerically via~\cite{nikitin2011fields},
\begin{equation}
\overleftrightarrow{G}_{R}^{s,p}(\mathbf{x}_{1},\mathbf{x}_{2},\omega)=\frac{ik_{\omega}}{8\pi}\int_{0}^{\infty}d q e^{ ik_{\omega}q_{z}(z_{2}+z_{1})}\overleftrightarrow{M}_{R}^{s,p},
\end{equation}
and
\begin{equation}
\overleftrightarrow{G}_{T}^{s,p}(\mathbf{x}_{1},\mathbf{x}_{2},\omega)=\frac{ik_{\omega}}{8\pi}\int_{0}^{\infty}dq e^{ ik_{\omega}[q_{z}z_{1}-q_{z}^{\prime}z_{2}]}\overleftrightarrow{M}_{R}^{s,p},
\end{equation}
where $k_{\omega}=\omega/c$ is the modular of the wave vector in free space, $q_{\alpha} =k_{\alpha}/k_{\omega},\alpha=x,y,z$ is the normalized dimensionless wave vector, $q=\sqrt{q_{x}^{2}+q_{y}^{2}}$ the projection of $\vec{q}$ on the $xy$-plane, and $q^{\prime}_z=\sqrt{\varepsilon(\omega)-q^2}$ with the relative permittivity of the outgoing media $\varepsilon(\omega)$. The kernals in the integrals are given by,
\begin{align}
\overleftrightarrow{M}_{R}^{s}
& =\frac{qr_{s}(q)}{q_{z}}\left[\begin{array}{ccc}
J_{0}+J_{2}\cos(2\phi_{0}) & J_{2}\sin(2\phi_{0}) & 0\\
J_{2}\sin(2\phi_{0}) & J_{0}-J_{2}\cos(2\phi_{0}) & 0\\
0 & 0 & 0
\end{array}\right],
\end{align}
\begin{align}
\overleftrightarrow{M}_{R}^{p} 
 & =-qr_{p}(q)\left[\begin{array}{ccc}
q_{z}\left[J_{0}-J_{2}\cos(2\phi_{0})\right] & -q_{z}J_{2}\sin(2\phi_{0}) & 2iqJ_{1}\cos\phi_{0}\\
-q_{z}J_{2}\sin(2\phi_{0}) & q_{z}\left[J_{0}+J_{2}\cos(2\phi_{0})\right] & 2iqJ_{1}\sin\phi_{0}\\
-2iqJ_{1}\cos\phi_{0} & -2iqJ_{1}\sin\phi_{0} & -2J_{0}q^{2}/q_{z}
\end{array}\right],
\end{align}
\begin{align}
\overleftrightarrow{M}_{T}^{s}
 & =\frac{qt_{s}(q)}{q_{z}}\left[\begin{array}{ccc}
J_{0}+J_{2}\cos(2\phi_{0}) & J_{2}\sin(2\phi_{0}) & 0\\
J_{2}\sin(2\phi_{0}) & J_{0}-J_{2}\cos(2\phi_{0}) & 0\\
0 & 0 & 0
\end{array}\right].
\end{align}
and
\begin{align}
\overleftrightarrow{M}_{T}^{p} 
 & =\frac{qt_{p}(q)}{q_{n}}\left[\begin{array}{ccc}
q_{z}^{\prime}\left[J_{0}-J_{2}\cos(2\phi_{0})\right] & -q_{z}^{\prime}J_{2}\sin(2\phi_{0}) & 2iq q_{z}^{\prime}J_{1}\cos\phi_{0}/q_{z}\\
-q_{z}^{\prime}J_{2}\sin(2\phi_{0}) & q_{z}^{\prime}\left[J_{0}+J_{2}\cos(2\phi_{0})\right] & 2iqq_{z}^{\prime}J_{1}\sin\phi_{0}/q_{z}\\
2iqJ_{1}\cos\phi_{0} & 2iqJ_{1}\sin\phi_{0} & 2J_{0}q^{2}/q_{z}
\end{array}\right].
\end{align}
Here, we have carried out the azimuth angle integral of $\vec{q}$ on the $xy$-plane and re-expressed the displacement $\mathbf{r}$ in the cylinder coordinate as $\mathbf{r}=r_{\perp}\mathbf{e}_{\rho}+z\mathbf{e}_{z}$
with $x=r_{\perp}\cos\phi_{0}$ and $\ y=r_{\perp}\sin\phi_{0}$. In these $M$-matrices, $J_{n}$ denotes Bessel function of $n$th order $J[n,q k_{\omega}r_{\perp}]$.

The Fresnel reflection and transmission coefficients of graphene-layer interface are given by~\cite{koppens2011graphene,nikitin2011fields}
\begin{equation}
r_{s}=\frac{q_{z}-q_{z}^{\prime}-2\alpha(\omega)}{q_{z}+q_{z}^{\prime}+2\alpha(\omega)},
\end{equation}
\begin{equation}
r_{p}=\frac{\varepsilon (\omega) q_{z}-q_{z}^{\prime}+2q_{z}q_{z}^{\prime}\alpha(\omega)}{q_{z}^{\prime}+\varepsilon(\omega)q_{z}+2q_{z}q_{2}^{\prime}\alpha(\omega)},
\end{equation}
\begin{align}
t_{s} & =1+r_{s},\\
t_{p} & =\frac{q_{1,z}}{q_{2,z}}\sqrt{\varepsilon(\omega)}(1-r_{p}),
\end{align}
where $\alpha(\omega)=2\pi\sigma(\omega)/\varepsilon_{0}c$ is the
dimensionless in-plane conductivity of the graphene. The optical conductivity of a graphene layer can be split into intra-band and inter-band contributions $\sigma(\omega)=\sigma_{{\rm intra}}(\omega)+\sigma_{{\rm inter}}(\omega)$ with~\cite{koppens2011graphene,biehs2013large}
\begin{align}
\sigma_{{\rm intra}}(\omega) & =\frac{2e^{2}k_{B}T}{\pi\hbar^{2}}\frac{i}{\omega+i/\tau_{D}}\log\left[2\cosh(E_{F}/2k_{B}T)\right],\\
 & \approx\frac{e^{2}}{\pi\hbar}\frac{iE_{F}/\hbar}{\omega+i/\tau_D}|_{T\rightarrow0}
\end{align}
and
\begin{align}
\sigma_{{\rm inter}}(\omega) & =\frac{e^{2}}{4\hbar}\left[H(\hbar\omega/2)+\frac{4i\hbar\omega}{\pi}\int_{0}^{\infty}dx\frac{H(x)-H(\hbar\omega/2)}{\hbar^{2}\omega^{2}-4x^{2}}\right]\\
 & \approx\frac{e^{2}}{4\hbar}\left[\Theta(\hbar\omega-2E_{F})+\frac{i}{\pi}\log\left|\frac{\hbar\omega-2E_{F}}{\hbar\omega+2E_{F}}\right|\right]|_{T\rightarrow0},
\end{align}
where $\tau_D$ is the relaxation time in the Drude model, $E_{F}$ the graphene's Fermi energy, and the function
\begin{align}
H(x) & =\frac{\sinh(x/k_{B}T)}{\cosh(E_{F}/k_{B}T)+\cosh(x/k_{B}T)}.
\end{align} 

\begin{figure}
\centering
\includegraphics[width=12cm]{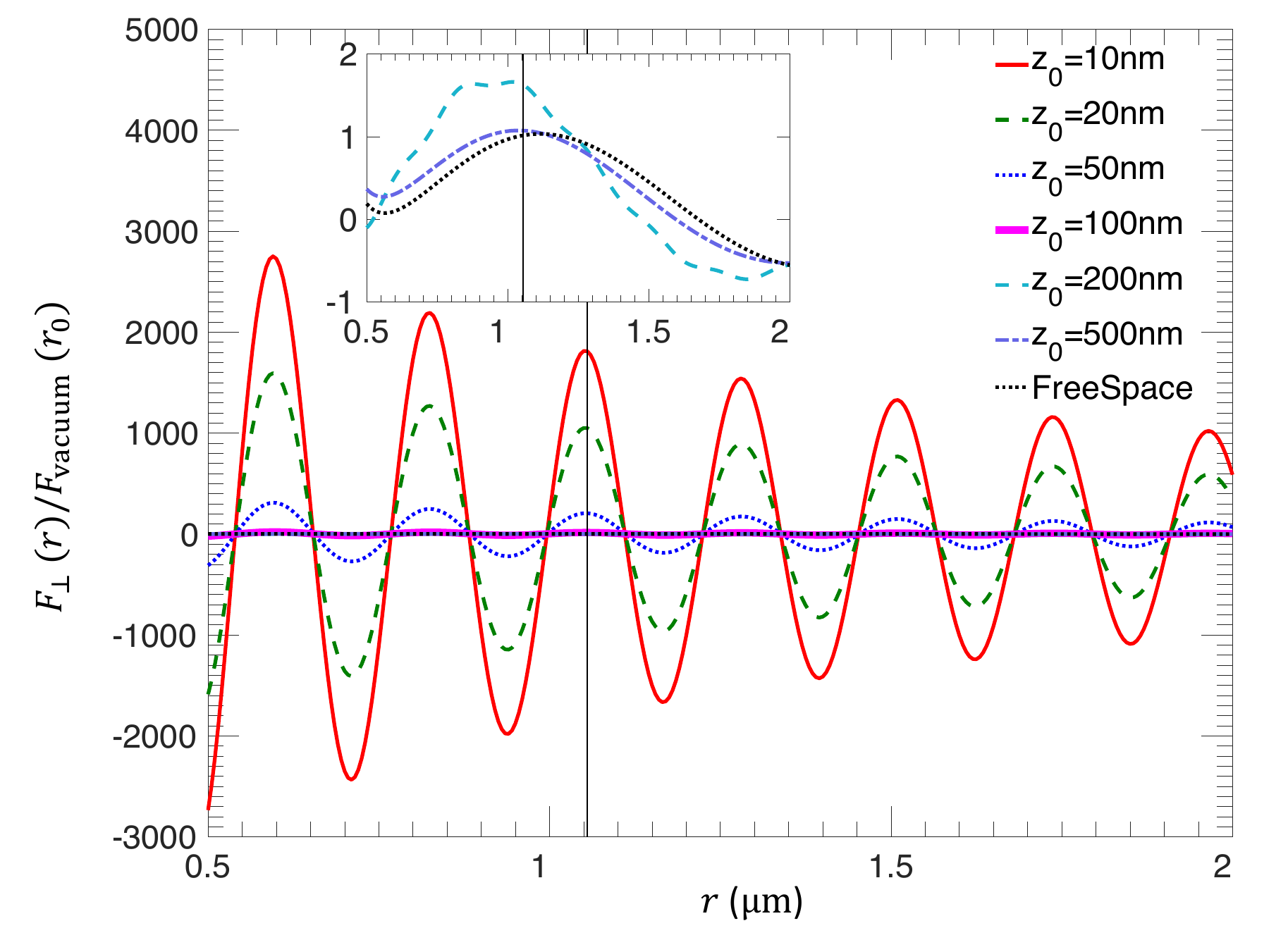}
\caption{\label{fig:graphene} The eigen value $F_{\perp}(r)$ of the RDDI force operator on state $|\Psi^{+}\rangle$ for two atoms on top of a graphene layer. Different curves denote different atom-interface distance $z_0$. In the subplot, we display the details of the curve for free-space case and the curves with $z_0 = 200$ nm and $z_0 = 500$ nm. Here, the electric dipole moments (along $z$-direction) of the atoms are perpendicular to the relative displacement $\mathbf{r}$ and $F_{\perp}(r)$ has been re-scaled by the eigen value $F_{\rm vacuum}(r_0)$ of the corresponding RDDI force operator in vacuum at $r_0 = 1.05 {\rm \mu m}$ (denoted by the vertical black line). The Fermi energy of the graphene is set as $E_{F} = 1.0$ eV and the relaxation time is taken as  $\tau_D = 10^{-13}$ s. To obtained a large enhancement in the RDDI force, the energy splitting of the two-level atoms is set as $\hbar\omega_0 = 0.7$ eV different from the optical transition in Rb atoms as shown in previous section. The graphene layer is considered to lie on an $\varepsilon(\omega_0) = 2.5$ substrate.  }
\end{figure}

The RDDI strength for two atoms on top of a graphene layer is given by
\begin{equation}
 U_{eg,ge}(r)=-\frac{\omega_{0}^{2}}{\varepsilon_{0}c^{2}}\boldsymbol{\mu}_i\cdot{\rm Re}\overleftrightarrow{G}(\mathbf{x}_{i},\mathbf{x}_{j},\omega_{0})\cdot\boldsymbol{\mu}_j  
\end{equation}
Then, the eigen value of the RDDI force operator on the state $|\Psi^{+}\rangle$ is obtained as $F(r) = -\partial U_{eg,ge}(r)/\partial r$. In Fig.~\ref{fig:graphene}, to show the enhancement in the RDDI force due to the graphene layer, we re-scale $F(r)$ with the eigen value $F_{\rm vacuum}(r_0)$ of the corresponding RDDI force operator in vacuum at $r_0 = 1.05 {\rm \mu m}$ (denoted by the vertical black line). Comparing with the subplot, we see that more than three order enhancement in the force can be obtained if the atoms are very close to the graphene layer ($z_0= 10$ nm). We also see that this enhancement decreases fast with the hight of the atoms $z_0$ and vanishes for $z_0 > 500$ nm.

In the main text, the corresponding time-dependent entanglement force induced by a single photon pulse has been displayed. The inter-atomic distance is set as $r=1.05 {\rm \mu m}$ as marked by the dark vertical line in Fig.~\ref{fig:graphene} and the atom-interface distance is set as $z_0 = 10, 20, 50$ nm. The pulse length $\tau_f$ has been optimized to get the maximum entanglement force as both the local spontaneous decay rate $\gamma_{ii}$ and the cooperative decay rates $\gamma_{ij}$ defined in Eq.~\ref{eq:gamma_ij} have also been greatly enhanced by the graphene layer.

\section{Time-dependent Master Equation for Quantum Pulse Scattering Processes\label{sec:TDMEQ}}

In this section, we study the dynamics of a two-level-atom pair. Different from the previous literatures,
we prepare the atom pair in the ground state $\left|gg\right\rangle $
instead of a single-excited state (e.g., $\left|eg\right\rangle $). In 2012, Ben \textit{et al}. derived a powerful
time-dependent master equation for $n$-photon broadband pulse interacting with an arbitrary quantum system. Here, we generalize this method to calculate the dynamical RDDI force. 



The total master equation including the single-photon pumping process
is given by, 
\begin{equation}
\frac{d}{dt}\tilde{\rho}(t)=[\hat{\hat{\mathcal{L}}}_{{\rm atom}}+\hat{\hat{\mathcal{L}}}_{{\rm pump}}]\tilde{\rho}(t),\label{eq:TDME}
\end{equation}
where $\tilde{\rho}(t)=\rho_{{\rm PN}}(t)\otimes\rho(t)$ is an effective
density matrix and we have introduced an extra qubit degree of freedom
$\rho_{{\rm PN}}(t)$ to characterize the photon number degree (see
more details in Ref.~\cite{Baragiola2012n-photon}). The initial
value of $\tilde{\rho}(t)$ is given by $\tilde{\rho}(0)=\hat{I}_{{\rm PN}}\otimes\rho(0)$,
where $\hat{I}_{{\rm PN}}$ is the two-dimensional identity matrix
and $\rho(0)=\left|gg\right\rangle \left\langle gg\right|$ is the
initial state of the atom pair. 

The the first term at right hand side (r.h.s.) of Eq.~(\ref{eq:TDME})
characterizes the free evolution of the atom pair without the pumping
\begin{align}
\hat{\hat{\mathcal{L}}}_{{\rm atom}}\tilde{\rho}(t) =& -i\left[\sum_{j}\omega_{0}\hat{\sigma}_{j}^{+}\hat{\sigma}_{j}^{-}+\sum_{i,j}\delta_{ij}\hat{\sigma}_{i}^{+}\hat{\sigma}_{j}^{-},\tilde{\rho}(t)\right]\nonumber \\
& +\sum_{ij}\frac{1}{2}\gamma_{ij}[2\hat{\sigma}_{i}^{-}\tilde{\rho}(t)\hat{\sigma}_{j}^{+}-\tilde{\rho}(t)\hat{\sigma}_{i}^{+}\hat{\sigma}_{j}^{-}-\hat{\sigma}_{i}^{+}\hat{\sigma}_{j}^{-}\tilde{\rho}(t)].
\end{align}
The second term characterizes the pumping of the single-photon pulse,
\begin{equation}
\mathcal{L}_{{\rm pump}}\tilde{\rho}=\sum_{j}\sqrt{\gamma_{0}}\eta_{j}\left\{ \xi(t-t_{j})[\hat{\tau}_{+}\rho_{{\rm tot}},\hat{\sigma}_{j+}]+\xi^{*}(t-t_{j})[\hat{\sigma}_{j-},\rho_{{\rm tot}}\hat{\tau}_{-}]\right\} ,\label{eq:pumping}
\end{equation}
with Pauli matrices $\hat{\tau}_{-}$ characterizing the absorption
of the single photon pulse. The parameter $\eta_{j}$ characterizes the pumping efficiency of the $j$th atom determined by its effective scattering cross section,  $t_{j}=(\mathbf{x}_{j}\cdot\vec{e}_{p})/c$ is the time of the center of the pulse arriving the $j$th atom, and 
\begin{equation}
\xi(t)=\frac{1}{\sqrt{2\pi}}\int_{0}^{\infty}d\omega\xi(\omega)e^{i\omega t},
\end{equation}
is the Fourier transform of the pulse spectrum function. For a Gaussian single photon pulse,
\begin{equation}
\xi(\omega)=\left(2\tau_{f}^{2}/\pi\right)^{1/4}\exp\left[-\tau_{f}^{2}(\omega-\omega_{0})^{2}\right],    
\end{equation}
its wave packet amplitude  in the time-space domain is given by,
\begin{equation}
\xi(t)=\left(\frac{1}{2\pi\tau_f^2}\right)^{1/4}\exp\left[-\frac{t^2}{4\tau_f^2}-i\omega_0 t \right].   
\end{equation}

In the main text, we assume the pulse propagates along the $x$-axis and arrives
at the two atoms at the same time $t_{1}=t_{2}=0$. The pumping efficiency
$\eta_{j}$ in practice shoule be much smaller than $1$~\cite{wang2011efficient,yang2018concept},
but its can be enhanced by adding a mode converter~\cite{sondermann2007design}.
In our simulation, we take $\eta_{1}=\eta_{2}=1/\sqrt{2}$ for the
homogeneous pumping case and $\eta_{1}=1,\,\eta_{2}=0$ for the locally
pumping case.

This effective master equation method can be straightforwardly generalized to $n$-photon Fock-State pulse case by
replacing the Pauli matrix $\hat{\tau}_{\pm}$ in Eqs.~(\ref{eq:TDME}-\ref{eq:pumping}) with
\begin{equation}
\hat{\tau}_{+}=\left[\begin{array}{ccccc}
0 & \sqrt{n} & 0 & 0 & 0\\
0 & 0 & \sqrt{n-1} & 0 & 0\\
0 & 0 & 0 & \ddots & 0\\
0 & 0 & \ddots & 0 & 1\\
0 & 0 & 0 & 0 & 0
\end{array}\right],\ \hat{\tau}_{-}=\left[\begin{array}{ccccc}
0 & 0 & 0 & 0 & 0\\
\sqrt{n} & 0 & 0 & 0 & 0\\
0 & \sqrt{n-1} & 0 & \ddots & 0\\
0 & 0 & \ddots & 0 & 0\\
0 & 0 & 0 & 1 & 0
\end{array}\right],
\end{equation}
and replacing the $2\times 2$ identity matrix $\hat{I}_{{\rm PN}}$ with the $(n+1)\times (n+1)$ identity matrix.

\begin{figure}
\centering
\includegraphics[width=14cm]{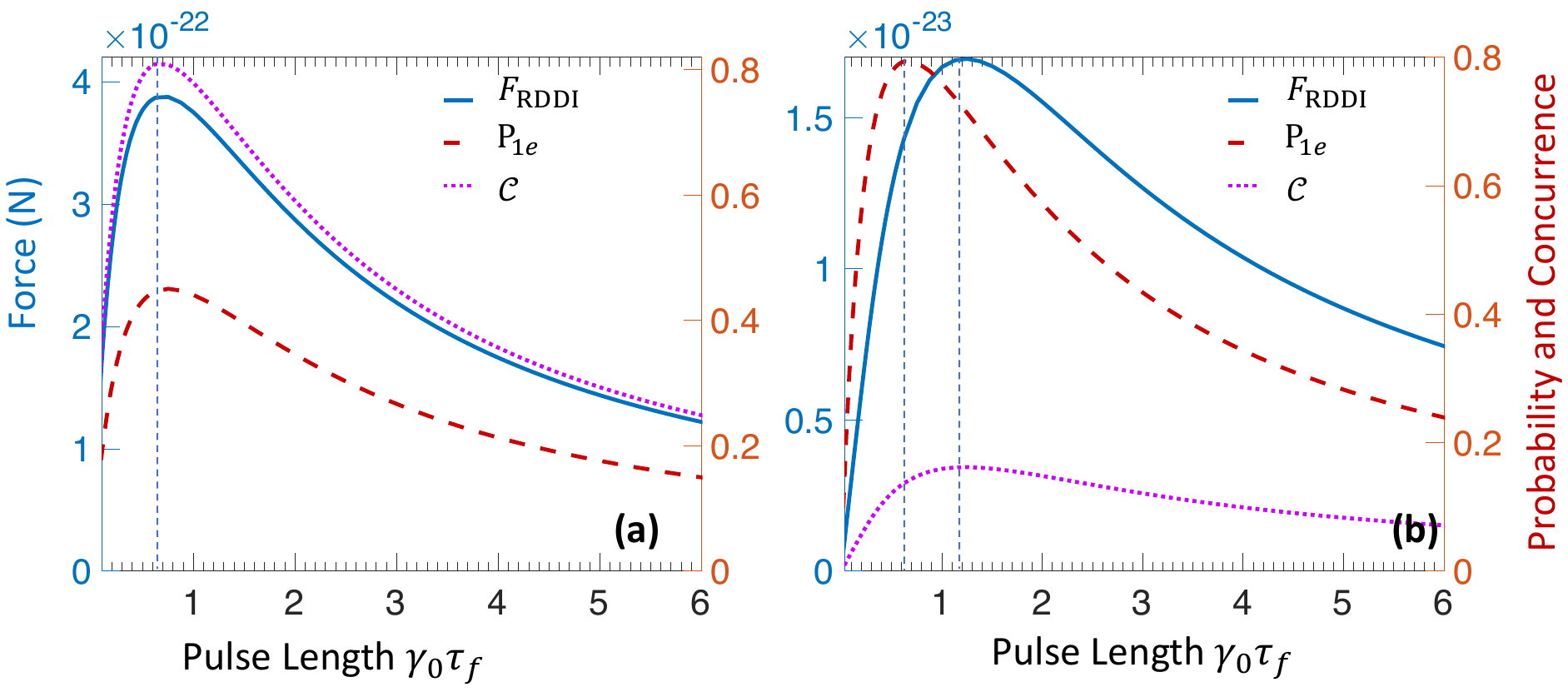}
\caption{\label{fig:S5}Optimization of the dynamic force $F_{{\rm \perp}}$
(the solid-blue line), concurrence $\mathcal{C}$ (the dashed-pink
line), and the excitation probability of the first atom $P_{1e}$
(the dotted-red line) via tuning the pulse length $\tau_{f}$. The
atom-atom distance is fixed at $r=1.2\,{\rm \mu m}$. (a) Homogeneous
pumping case. (b) Local pumping case.}
\end{figure}

Actually, $\tilde{\rho}(t)$ is not a real density matrix of a physical
system, as ${\rm Tr}\tilde{\rho}(0)= n$ for $n$-photon Fock-state pulse. Thus, only its projection
on the specific subspace has physical meaning. The expected value
of any atomic operator $\hat{O}$ is given by
\begin{equation}
\langle\hat{O}\rangle_{t}\equiv {\rm Tr}[\hat{O}\rho(t)] = {\rm Tr}[\tilde{\rho}(t)\left(\hat{P}\otimes\hat{O}\right)],
\end{equation}
where $\hat{P}$ is the projection operator of the extra
qubit degree with the only non-zero element $P_{11}=1$. We also note that, to handle the coherent-state pulse case, we only need to replace all the photon related operators (i.e., $\hat{\tau}_{\pm}$, $\hat{I}_{\rm PN}$, and $\hat{P}$) with the constant $1$. This powerful time-dependent master equation (\ref{eq:TDME}) can be used to uniformly study the quantum photon pulse scattering process. 


We can also enhance the RDDI force by changing the pulse length $\tau_{f}$ to optimize the two-body entanglement (see Fig.~\ref{fig:S5}). Here, we see that, for homogeneous pumping case with $\eta_1=\eta_2=1/\sqrt{2}$, the optimal pulse length maximizes the local excitation probability of the first atom $P_{1e}$, the inter-atomic force $F_{\rm RDDI}$, and the concurrence $\mathcal{C}$ simultaneously {[}see Fig.~\ref{fig:S5}(a){]}. But, for local pumping
case with $\eta_1=1$ and $\eta_2=0$, only the pulse length optimizing $\mathcal{C}$ maximizes the RDDI force {[}see Fig.~\ref{fig:S5}(b){]}. A shorter pulse optimizes the photon absorption probability $P_{1e}$, but the entanglement and the force are suppressed due to the low entanglement generation rate via the weak RDDI coupling and the fast spontaneous decay rates of the atoms. Thus, the homogeneous pumping is a more efficient way to generate the entanglement force.  

\bibliography{main}
\bibliographystyle{unsrt}
\end{document}